\documentclass[preprint,review,3p]{elsarticle}
\usepackage{natbib}

\newcommand{\cb}[1]{\ifmmode {\boldsymbol{#1}}\else ${\boldsymbol{#1}}$\fi}
\newcommand{\cp}[1]{\ifmmode {\mathcal{#1}}\else ${\mathcal{#1}}$\fi}

\usepackage[T1]{fontenc}
\usepackage{psfrag,epsfig,graphics}
\usepackage{psfrag}
\usepackage{multirow}
\usepackage{multicol}
\usepackage{color}
\usepackage{colortbl}
\usepackage[centertags]{amsmath}
\usepackage{amsfonts}
\usepackage{amssymb}
\usepackage{newlfont}
\usepackage{graphicx}
\usepackage{amsmath}
\usepackage{tabularx}
\usepackage{float}
\usepackage{comment}
\usepackage{gensymb}
\usepackage{soul}
\usepackage{colortbl}
\definecolor{Gray}{gray}{0.9}
\newcolumntype{a}{>{\columncolor{Gray}}c}

\usepackage[ruled]{algorithm2e}
\usepackage{subfigure}
\usepackage{lineno}
\usepackage{svg}
\usepackage{longtable}
\usepackage{lscape}
\usepackage{multirow}
\usepackage{rotating}
\usepackage{graphicx}
\usepackage[colorlinks]{hyperref}
\usepackage[withpage]{acronym}
\usepackage{color,soul}
\definecolor{Gray}{gray}{0.9}

\soulregister\cite7
\soulregister\ref7
\soulregister\pageref7

\newcolumntype{R}[1]{>{\RaggedRight}p{#1}}

\usepackage{mathrsfs}
 \makeatletter
    \def\ps@pprintTitle{%
       \let\@oddhead\@empty
       \let\@evenhead\@empty
       \let\@oddfoot\@empty
       \let\@evenfoot\@oddfoot
    }
    \makeatother

\usepackage{setspace}
\journal{}

\begin{document}

\begin{frontmatter}

\title{Towards Zero Trust Security in Connected Vehicles: A Comprehensive Survey} 

\author[a]{Malak Annabi}
\author[a]{Abdelhafid Zeroual}
\author[b]{Nadhir Messai}

\address[a]{Laboratoire de Recherche d’Electronique de Skikda LRES, Faculté de technologie, Université 20 Août 1955-Skikda, Skikda, Algeria}
\address[b]{Université de Reims Champagne Ardenne, CReSTIC EA 3804, 51097 Reims, France}

\begin{abstract}
\doublespacing

 Zero Trust is the new cybersecurity model that challenges the traditional one by promoting continuous verification of users, devices, and applications, whatever their position or origin. This model is critical for reducing the attack surface and preventing lateral movement without relying on implicit trust. Adopting the zero trust principle in Intelligent Transportation Systems (ITS), especially in the context of connected vehicles (CVs), presents an adequate solution in the face of increasing cyber threats, thereby strengthening the ITS environment.
This paper offers an understanding of Zero Trust security through a comprehensive review of existing literature, principles, and challenges. It specifically examines its applications in emerging technologies, particularly within connected vehicles, addressing potential issues and cyber threats faced by CVs. Inclusion/exclusion criteria for the systematic literature review were planned alongside a bibliometric analysis. Moreover, keyword co-occurrence analysis was done, which indicates trends and general themes for the Zero Trust model, Zero Trust implementation, and Zero Trust application. Furthermore, the paper explores various ZT \textcolor{red}{\hl{models}} proposed in the literature for connected vehicles, shedding light on the challenges associated with their integration into CV systems.  Future directions of this research will focus on incorporating Zero Trust principles within Vehicle-to-Vehicle (V2V) and Vehicle-to-Infrastructure (V2I) communication paradigms. This initiative intends to enhance the security posture and safety protocols within interconnected vehicular networks. The proposed research seeks to address the unique cybersecurity vulnerabilities inherent in the highly dynamic nature of vehicular communication systems.


\end{abstract}

\end{frontmatter}
\section{Introduction}
\label{sect: introduction}

The evolving growth of the interconnected world and the adoption of advanced technologies like cloud computing, the Internet of Things (IoT), and Artificial Intelligence (AI) permit seamless communication and innovation. This interconnectivity has resulted in cyber security risks that include data breaches and cyber-attacks. 
The traditional perimeter security approaches need to be improved in the face of these complex threats. Zero Trust offers enhanced visibility, data protection, and defence against growing cyber threats by continuous authentication and verification of every entity on the network \cite{shen2024endpoint}. Also, it helps save money by combining tools and working more efficiently, which makes it a great model for keeping the network secure \cite{ACT-IAC:2019:Online}.

To ensure a methodologically sound and transparent systematic literature review, a comprehensive inclusion and exclusion criteria framework was formulated and presented in Table \ref{tab:table1}. This methodological approach aligns with established guidelines for conducting systematic reviews in scientific research, providing a standardized protocol for selecting and filtering relevant scholarly works \cite{moher2009preferred}. Implementing such criteria ensures a rigorous selection process for relevant literature, thereby enhancing the validity and reliability of the review's findings. This approach also mitigates selection bias and enhances the overall quality of the review.

\begin{longtable}[c]{|m{2.5cm}|m{6.25cm}|m{6.25cm}|}

\caption{Inclusion / Exclusion criteria for the systematic review}    
\label{tab:table1}\\
\hline
    \textbf{Criteria}  &  \textbf{Inclusion} &  \textbf{Exclusion}\\
\hline
Time frame & From $2018$ to $2024$ & Prior $2018$ \\
  \hline
  Language & Published in English &  Published in other languages \\
  \hline 
  Type & Peer-reviewed, review, articles, conference papers & Editorial, note, letters \\
  \hline 
  Topic & Studies specifically addressing zero trust architecture or principles & Studies addressing other types of cybersecurity models \\
  \hline
  Comparative analysis & Surveys that explore the theoretical foundations, implementation strategies, or adoption trends of zero trust & Other surveys not discussing zero trust model \\
  \hline 
  Applications & Research focusing on the application of zero trust principles in cloud computing, IoT, and 5G/6G technologies & Studies not specifically addressing zero trust principles or applications in the respective domains\\
  \hline 
  Connected vehicles & Studies focusing on zero trust principles applied in connected vehicle environments & Studies focusing solely on vehicle security without the application of zero trust\\
  \hline  

\end{longtable}
As shown in Table \ref{tab:table1}, the systematic review employed a stringent set of inclusion and exclusion criteria to ensure methodological rigour and relevance. The temporal scope was constrained to publications between 2018 and 2024, specifically focusing on zero trust architecture (ZTA). This timeframe aligns with the rapid evolution of ZTA concepts and implementations. The initial analysis phase encompassed surveys highlighting the theoretical foundations, implementation strategies, and adoption trends of ZTA. Additionally, research exploring the application of zero trust principles in cloud computing, Internet of Things (IoT), and 5G/6G technologies was included, reflecting the growing integration of ZTA across diverse technological domains.
The primary emphasis of this study, however, was on investigations that leveraged ZTA principles to improve the cybersecurity of connected vehicles. This focus reflects the increasing criticality of securing automotive systems in the era of intelligent transportation.
Studies pertaining to alternative cybersecurity models, those published before 2018, and those in languages other than English were excluded from the review. To ensure scholarly rigour, the analysis was restricted to peer-reviewed articles, systematic reviews, and conference papers. Editorials, notes, and letters were omitted due to their typically lower level of empirical content.
The subsequent sections of this paper elucidate the findings derived from this systematic analysis, providing a comprehensive overview of the current state of ZTA in connected vehicle cybersecurity.

\begin{figure}[H]
  \centering
   \begin{minipage}[b]{0.49\textwidth}
    \includegraphics[width=1.03\textwidth]{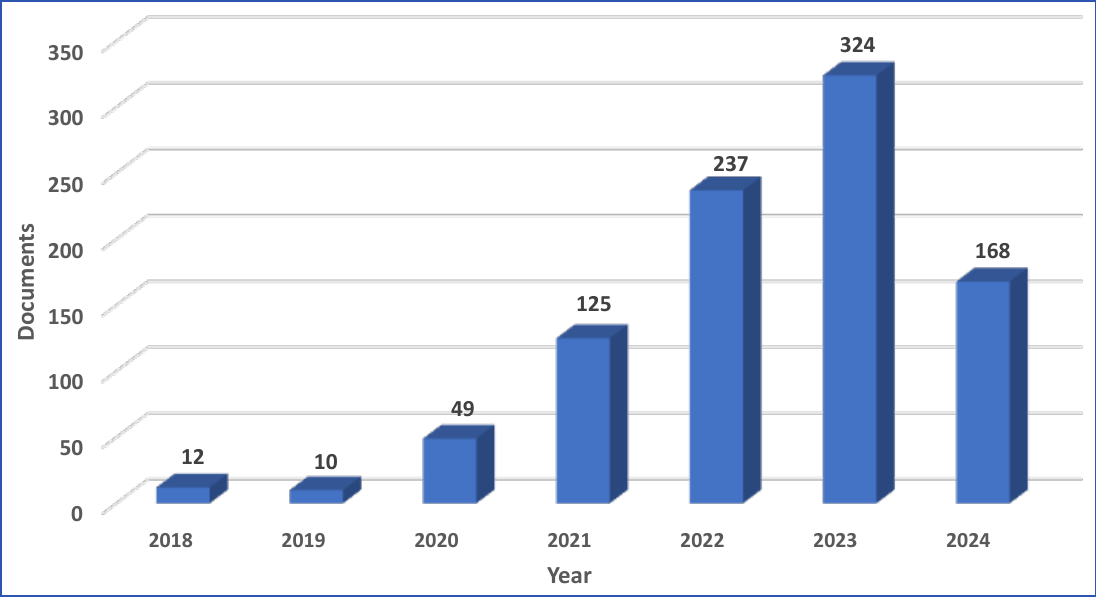}
    \caption{Evolution of the Zero Trust literature (Scopus database)}
    \label{fig:general_hist}
\end{minipage}
    \hfill
  \centering
  \begin{minipage}[b]{0.49\textwidth}
    \includegraphics[width=1.03\textwidth]{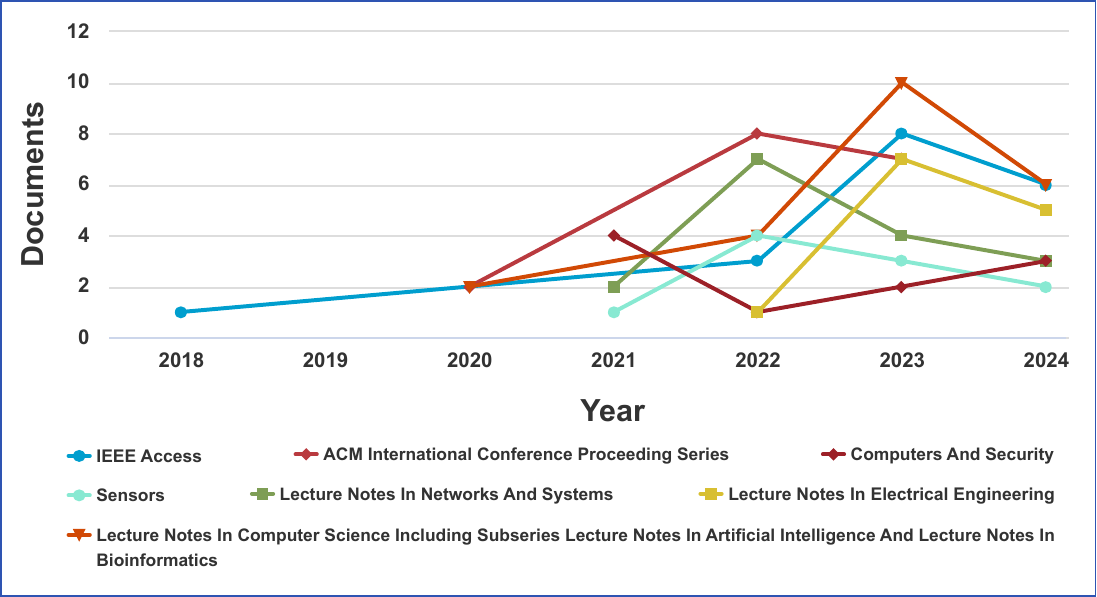}
    \caption{Evolution of the Zero Trust literature based on Publishing Sources}
    \label{fig:journal_hist1}
    \end{minipage}
    \end{figure}

The bibliometric analysis of zero trust literature, based on data from the Scopus database, provides a comprehensive overview of the academic aspect, defining the cybersecurity evolution of the research and highlighting key features. The research methodology conducted to extract from the title, keyword, and abstract was: TITLE-ABS-KEY((zero AND trust AND architecture) OR (Zero Trust) OR (Zero Trust-architecture)) AND PUBYEAR $ > $ 2017. In total, there have been 8 documents in the last ten years. Notably, there has been a significant surge in scholarly output from 2018 to 2024 , as evidenced by a rising number of publications. Specifically, from 2021 to 2024, the number of papers increased steadily, with 125, 237, 324, and 168 papers published, respectively, during these years. Figure \ref{fig:general_hist} illustrates this evolution, highlighting the growing significance and attention devoted to ZT within the cybersecurity domain. Figure \ref{fig:journal_hist1} further illustrates our literature analysis, which is based on published sources.\\

\begin{figure}[H]
  \centering
  \begin{minipage}[b]{0.49\textwidth}
    \includegraphics[width=1\textwidth]{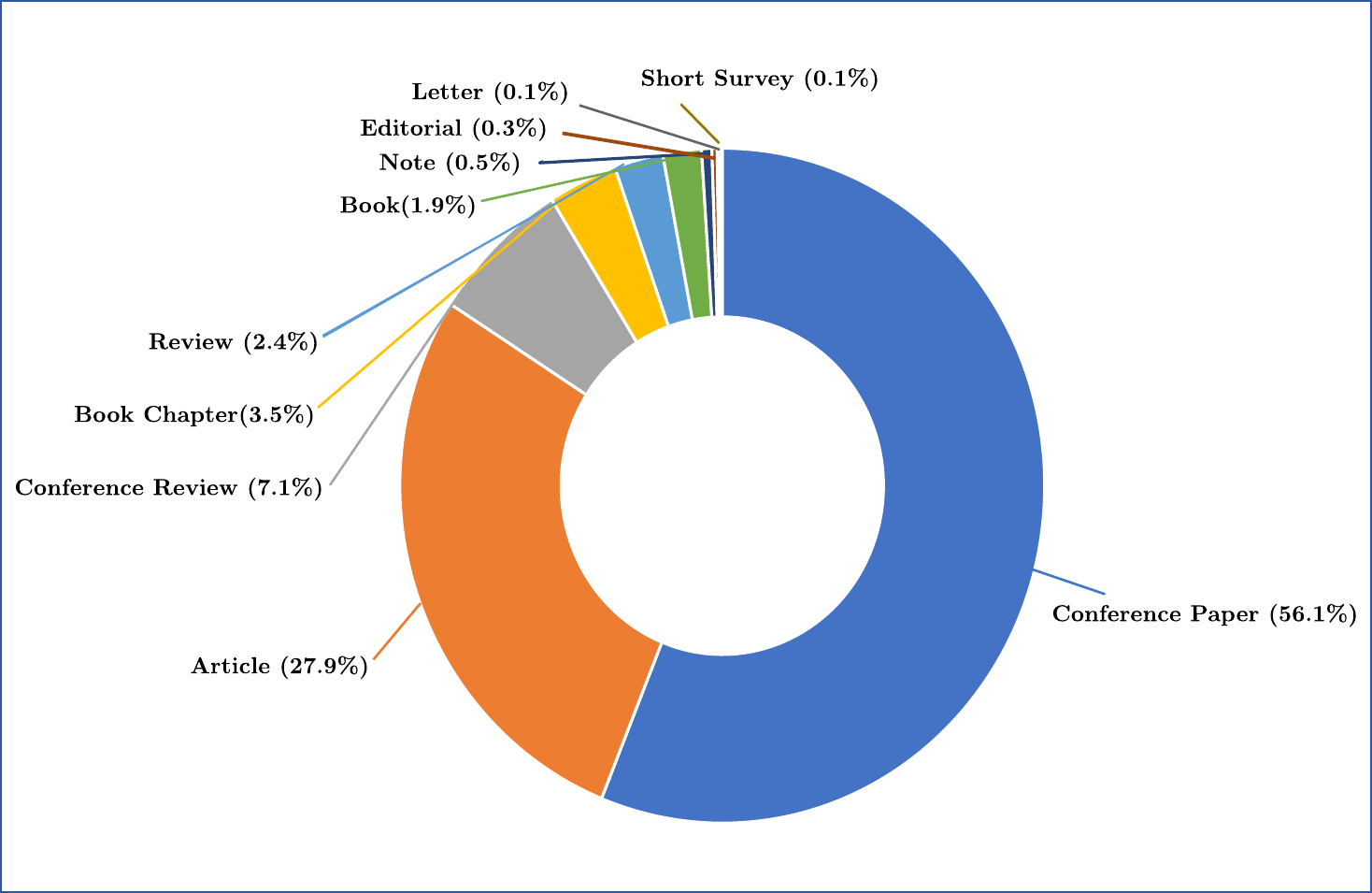}
    \caption{Classification of the documents based on their type (Scopus Database)}
    \label{fig:journal_hist3}
    \end{minipage}
    \hfill
  \centering
  \begin{minipage}[b]{0.49\textwidth}
    \includegraphics[width=1\textwidth]{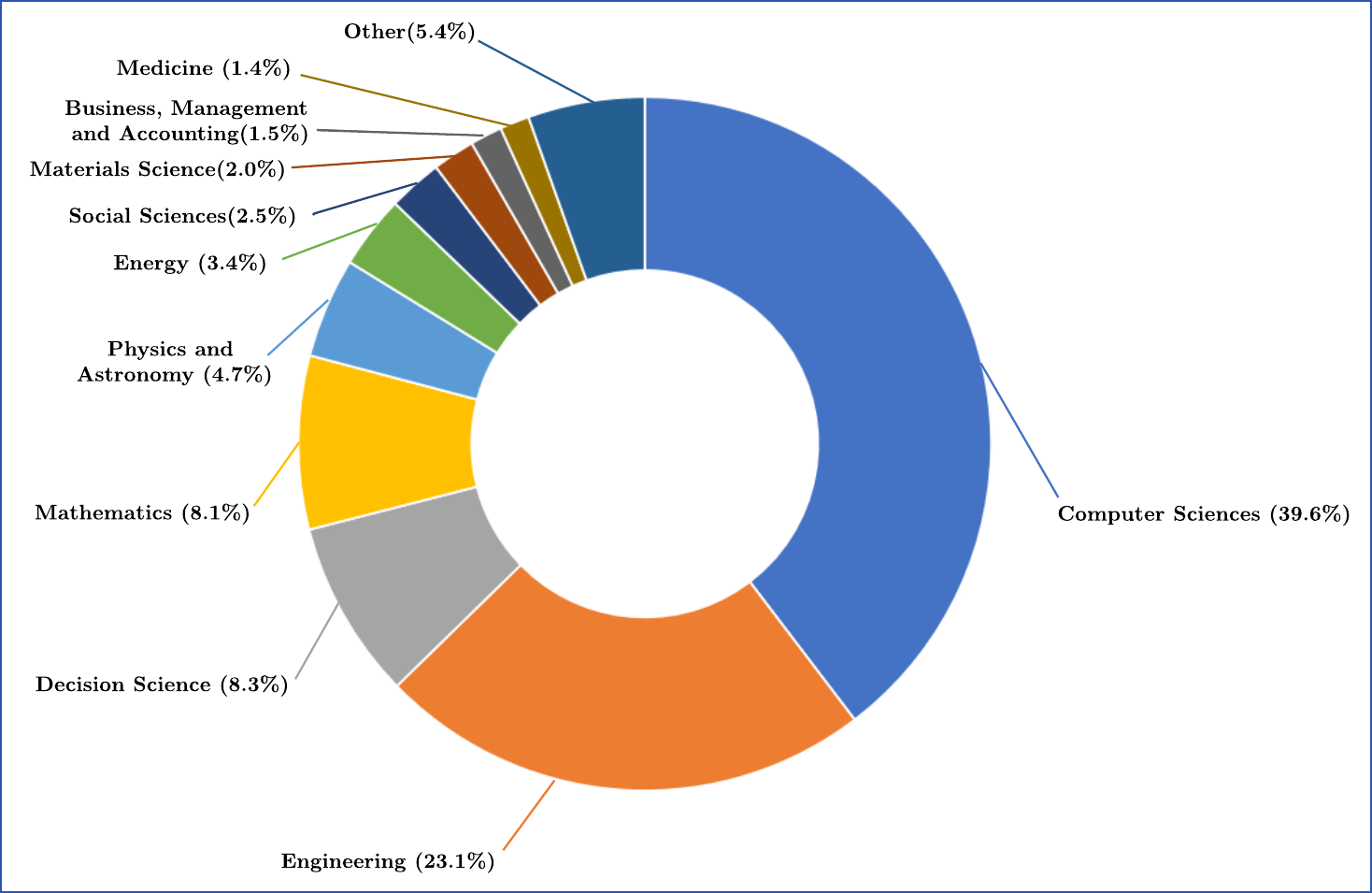}
    \caption{Classification of the documents based on their subject area (Scopus Database)}
    \label{fig:journal_hist4}
    \end{minipage}
\end{figure}

 These documents are categorized into two main categories, document types and subject areas. Analyzing the document types reveals that conference papers are the most prevalent, constituting 56.1\% of the total documents, followed by articles at 27.9\%. Conference Reviews, Book Chapters, Reviews, Books, and other types, as shown in Figure \ref{fig:journal_hist3}, make up a smaller percentage and indicate a varied composition of scholarly contributions. The documents encompass a broad spectrum of subject areas, as illustrated in Figure \ref{fig:journal_hist4}. Computer Science is the most prominent field at 39.6\%, with engineering closely behind at 23.1\%. Decision Sciences, Mathematics, Physics, and Astronomy exhibit sustainable representation, accounting for 8.3\%, 8.1\%, and 4.7\% of the documents. Energy, Social Sciences, Materials Science, Business, Management, Accounting, Medicine, and other areas contribute to the diverse range of research interests represented in the dataset extracted from the Scopus database. These visual representations offer valuable insights into the ZT landscape, including trends, key contributors, and pivotal works. Notably, the NIST special publication \cite{rose2020nist}, has significantly influenced numerous researchers in shaping their work.\\
 
\subsection{Keyword co-occurrence analysis}
Keywords in research papers provide crucial information about the content and significance of the paper. Analyzing keyword co-occurrences helps uncover trends and research variations within specific fields. This analysis is a common method in scientometrics, revealing the structure of the academic domains and highlighting research frontiers\cite{shi2019visualization}.\\

 \begin{figure}[H]
 \centering
   \includegraphics[width=0.8\textwidth]{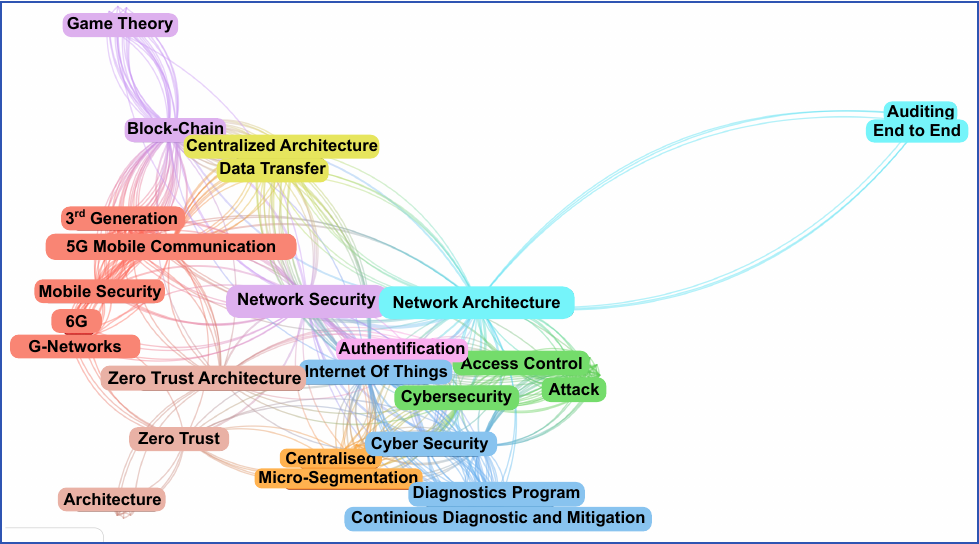}
    \caption{Network keyword co-occurrence with VOSviewer}
    \label{fig:network vosviewer5}
    \end{figure}
    
\begin{figure}[H]
  \centering
    \includegraphics[width=0.9\textwidth]{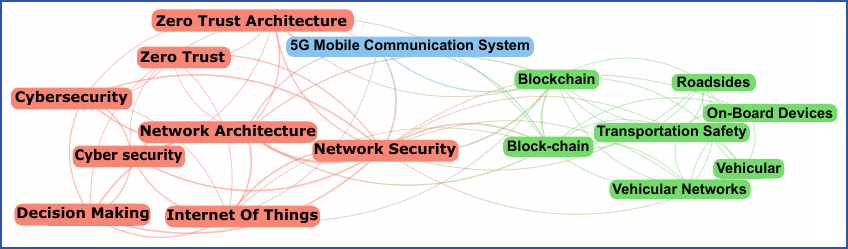}
    \caption{Network map of the high occurrence keywords alongside the ones related to the vehicles}
    \label{fig:14 keywords6}
\end{figure}

VOSviewer is used to analyze the network of keyword co-occurrences in the ZT field. Across all chosen articles in the datasets, 139 keywords meet the minimum threshold of one occurrence, indicating that each keyword appears at least once. The network of these co-occurrences is shown in Figure \ref{fig:network vosviewer5}. Keywords are represented as nodes, and the size of each one corresponds to the frequency of the keyword occurrence. Links between nodes signify relationships between the keywords they represent. The closeness of two nodes indicates a higher frequency of those keywords appearing together. In addition, the colour of each node indicates the cluster it belongs to; as shown in Figure \ref{fig:network vosviewer5}, there are 9 clusters. The most significant node corresponds to "Network Security" with an occurrence of 8 and total link strength of 140, followed by "Network Architecture" with 7 occurrences and 124 of strength, "5G mobile communication systems", "Internet of Things", "block-chain", alongside the keywords related to the ITS as illustrated in Table \ref{tab1}, "on-board devices", "roadsides", "transportation safety", "vehicles", and "vehicular networks" with an occurrence of 1 and total link strength of 14. Figure \ref{fig:14 keywords6} displays the network map containing the keywords summarized in Table \ref{tab1}. The first 11 keywords meet the minimum threshold of appearing at least two times. \\

 \begin{table}  
 \begin{center}
 \caption{High Keywords occurrences}\label{tab1}
  \begin{tabular}{|c|c|c|} 
 \hline
   Keyword   & Occurrences & Total link of strength \\
   \hline
     network security & 8 & 140 \\
     \hline
     network architecture & 7 & 124 \\
     \hline
     5g mobile communication systems & 3 & 56 \\
     \hline 
     Internet of Things & 3 & 56 \\
     \hline
     block-chain & 3 & 55 \\
     \hline 
     blockchain & 2 & 55\\
     \hline
     decision making & 3 & 54 \\
     \hline
     cyber security & 3 & 53 \\
     \hline
     cybersecurity & 3 & 52 \\
     \hline
     zero trust architecture & 3 & 48 \\
     \hline
    zero trust & 3 & 45 \\
     \hline
    on-board devices & 1 & 14 \\
     \hline
   roadsides& 1 & 14 \\
     \hline
     transportation safety & 1 & 14 \\
     \hline
     vehicles & 1 & 14 \\
     \hline
      vehicular networks & 1 & 14 \\
     \hline
\end{tabular}
 \end{center}
 \end{table}

Figure \ref{fig:network vosviewer5}, Figure \mbox{\ref{fig:14 keywords6}} and Table \mbox{\ref{tab1}} collectively clarify a pertinent correlation, allowing us to comprehensively analyse Zero Trust's prevalence across diverse research domains. This multifaceted examination enables us to delineate the Zero Trust model's relationship with the spectrum of scientific terminologies, each representing distinct applications and contexts of the model. The graphical representation in Figure \ref{fig:network vosviewer5} and  Figure \ref{fig:14 keywords6} provides a visual synthesis of these interconnections, illustrating the semantic and conceptual linkages between the Zero Trust Model and the other terminologies enumerated in Table \ref{tab1}. These connections are derived from an extensive corpus of scientific literature indexed in the Scopus database, offering a robust foundation for our analysis. The synergistic examination of these representations offers a nuanced perspective on the pervasive influence of the Zero Trust Model across the scientific landscape. This comprehensive analysis not only elucidates the current state of Zero Trust research but also potentially identifies emerging trends and unexplored avenues for future investigation. The diverse array of associated terms underscores the model's adaptability and broad relevance, suggesting its continued significance in addressing security challenges across multiple domains.

\subsection{Comparative Analysis of Existing Surveys} 
Numerous studies have investigated the architectural and technical elements of ZTA. In this part, we provide a brief literature review of existing surveys on ZTA, highlighting the unique contributions of each, as summarized in Table 3. Authors in \cite{buck2021never} conducted a survey that scrutinized ZTA papers employing a search model that differentiated academic literature from grey literature (non-academic, commercial, or private sources.) and examined the drawbacks and expenses of ZTA considering the economic impact and user viewpoint within a blockchain context. Furthermore, researchers in \cite{he2022survey} provide a research investigation into the pros and cons of access control models and authentication protocols while contrasting prevalent assessment techniques used to evaluate trust. Similar to \textit{Syed et al.}\cite{syed2022zero} exploring the latest authentication and access control technologies within various ZTA contexts, focusing on ZTA encryption, micro-segmentation, and security automation, and expanding its scope to encompass software-defined perimeters and challenges. \textit{Pittman et al.}\cite{pittman2022towards} propose implementing Zero Trust tenets directly to data objects rather than access pathways. 
In \cite{cao2024automation}, researchers centre on employing AI techniques to automate and orchestrate ZTA, highlighting their significant potential and roles in addressing the challenges and enabling automation and orchestration of the model.
\textit{Teerakanok et al.} \cite{teerakanok2021migrating} explores the obstacles, necessary procedures, and relevant factors involved in transitioning from traditional architecture to ZTA.

Surveys historically emphasized ZTA architecture development and management, while current ones focus on comprehensive evaluations of theoretical models and application scenarios within ZTA. According to the literature, none of the surveys, to our knowledge, handle ZTA technologies within the context of connected vehicles.



\begin{longtable}[c]{|m{3.65cm}|m{4cm}|m{0.9cm}|m{0.9cm}|m{0.9cm}|m{0.9cm}|m{0.9cm}|m{0.7cm}|}

\caption{Analysis of existing surveys on Zero Trust \\ (\color{blue} {\checkmark} \color{black} : Discussed  $-$: Partially Mentioned  \color{red} X  \color{black} : Not Mentioned )} \\
\hline
\centering \textbf{Reference} & \centering \textbf{Contributions} & \rotatebox[origin=c]{90}{\parbox{17mm}{\centering \textbf{ZT} Principles}} &  \rotatebox[origin=c]{90}{\parbox{20mm}{\centering \textbf{ZT} Architecture}} & \rotatebox[origin=c]{90}{\parbox{25mm}{\centering \textbf{ZT} Maturity Model}} &  \rotatebox[origin=c]{90}{\parbox{20mm}{\centering \textbf{ZTA}  Challenges}} & \rotatebox[origin=c]{90}{\parbox{20mm}{\centering \textbf{ZTA} applications}} &  \rotatebox[origin=c]{90}{\parbox{20mm}{\centering \textbf{ZTA}-CVs }} \\
\hline
\textit{ Buck et al.} \cite{buck2021never} & Synthesizing Zero Trust Foundations and Identifying Knowledge Gaps across Industry and Academia & \color{blue} \checkmark & \color{blue} \checkmark & \color{red} X & \color{red}X & \color{red} X & \color{red} X \\ 
\hline
\textit{ He et al.}\cite{he2022survey} & Analysing ZTA core technologies and comparing their advantages and drawbacks &  \color{blue} \checkmark &  \color{blue} \checkmark & \color{red}X & \color{red} X & \color{red} X & \color{red} X \\
\hline
\textit{ Syed et al.}\cite{syed2022zero} & Exploring Access Control and Authentication of ZTA  &  \color{blue} \checkmark & \color{blue} \checkmark & \color{red}X &  \color{blue} \checkmark & \color{red}X & \color{red}X \\
\hline
\textit{ Pittman et al.} \cite{pittman2022towards} & Adopting Zero Trust Principles for Safeguarding Data Objects &  \color{blue} \checkmark &  $-$ &\color{red}X & \color{red}X & \color{red}X & \color{red}X \\
\hline
\textit{ Sarkar et al.}\cite{sarkar2022security} & Grouping and Comparing New Elements in Zero Trust Models for Cloud Networks & \color{blue} \checkmark &  \color{red}X &\color{blue}\checkmark & \color{blue}\checkmark & $-$ & \color{red}X \\
\hline
\textit{ Alevizos et al.} \cite{alevizos2022augmenting} & Examining ZTA-based models and blockchain-driven intrusion detection and prevention to strengthen endpoint security & \color{blue}\checkmark & \color{blue}\checkmark & \color{red}X & \color{red}X & \color{red}X & \color{red}X \\
\hline
\textit{ Cao et al.}\cite{cao2024automation}& Evaluating AI-centered methods for addressing the technical aspects of automating ZTA & \color{red}X & \color{blue}\checkmark & \color{blue}\checkmark & \color{blue}\checkmark & \color{blue}\checkmark &  {$-$}
\\
\hline
 \textit{ Teerakanok et al.} \cite{teerakanok2021migrating} & Examining the challenges and essential aspects of transitioning to ZTA   & \color{blue}\checkmark& \color{blue}\checkmark & \color{red}X & \color{red} X & \color{red}X & \color{red} X \\ 
 \hline
\textit{ Yan et al.} \cite{yan2020survey} & Evaluating the fundamental technologies within ZTA components & \color{blue}\checkmark & \color{blue}\checkmark & \color{red} X & \color{red}X &  \color{blue}\checkmark  & \color{red} X \\
\hline 
\textit{ Hongzhaoning et al.}\cite{kang2023theory} & Introducing the "trust base" and outlining future research trends& \color{blue}\checkmark & \color{blue}\checkmark & \color{red}X & \color{blue} \checkmark & \color{blue} \checkmark &  \color{red}X  \\
\hline
\textit{ This Survey.} & Exploring the Zero Trust model and its application in enhancing cybersecurity, particularly in connected vehicles (CVs) &\color{blue}\checkmark & \color{blue}\checkmark & \color{blue} \checkmark &\color{blue}  \checkmark &\color{blue}  \checkmark & \color{blue}  \checkmark \\
\hline
\end{longtable}


As the automotive industry embraces digital transformation and connectivity, the security challenges associated with connected vehicles become increasingly prominent. Our motivation for linking the zero trust principle with transportation stems from the critical challenges that threats and attacks have increasingly targeted connected vehicles. Existing literature has proposed various techniques to enhance their security, including blockchain and machine learning. Various studies have explored the application of ZT across different domains, such as the Cloud, IoT, and 5G \ 6G. There is a significant gap in the literature regarding its specific implications and benefits in the realm of connected vehicles (CVs).  Building upon these foundations and being inspired by adopting the zero trust model in other sectors. This paper addresses the pressing need for a comprehensive survey on the zero trust security paradigm in light of its critical importance in ITS, particularly in the connected vehicle landscape.  The contribution of this paper is to explore the ZT security paradigm by offering a holistic understanding of its principles and applications, showcasing the adaptability of ZT in diverse technological landscapes. Focusing on connected vehicles, the paper addresses unique security concerns, emphasizing the role of ZT in enhancing automotive security. Furthermore, the paper offers a comprehensive analysis of ZT technology within the context of connected vehicles, contributing to existing surveys.


\label{sect: Survey's Structure}
\begin{figure}[H]
  \centering
\includegraphics[width=0.8\textwidth]{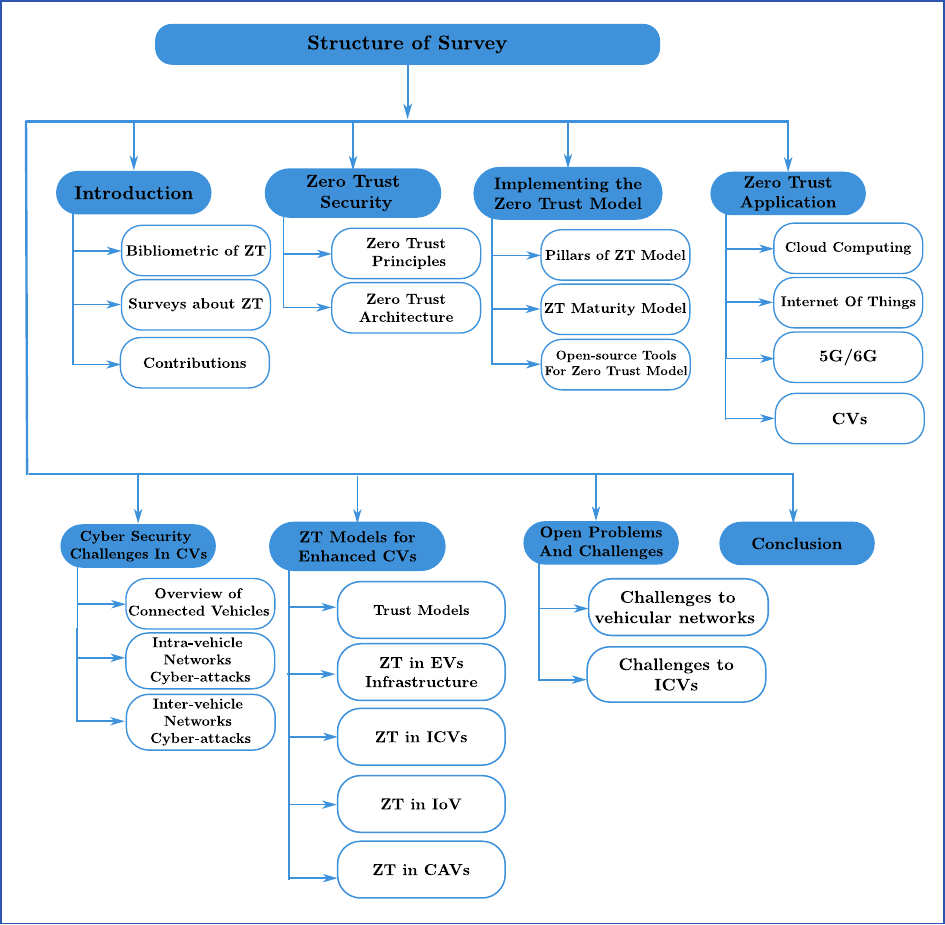}
\caption{Structure of The Survey}
    \label{fig:structure}
\end{figure}

\noindent This paper, as illustrated in Figure \ref{fig:structure}, is organized as follows:

Section 1 presents an introduction with bibliometric analysis of existing literature and surveys on zero trust security. Section 2 identifies the ZT concept, principles, and architecture. Section 3 introduces the Zero Trust implementation process. ZT applications across different fields, Cloud, IoT, 5G/6G, and connected vehicles are presented in Section 4. Section 5 provides insights into the communication system of CVs, the cyber-attacks, and the threats they face. Section 6 examines the various approaches to zero trust within the context of connected vehicles, along with the trust models. Section 7 discusses open problems and challenges. Finally, Section 8 concludes the survey.

\section{Zero Trust Generalities}

The Zero Trust model is a cybersecurity philosophy founded on "Never Trust, Always Verify." In 2004, the Jericho Forum, a collective of Chief Information Security Officers (CISOs) based in the United Kingdom \cite{d2021building, rose2020nist}, introduced the idea of deperimeterization. And then, in 2010, it gained interest from John Kindervag \cite{kindervag2010build, cao2024automation, he2022survey}, a former Forrester analyst. The idea challenges the traditional security model by assuming that all users, devices, and applications, regardless of location within or outside the network, are potential security threats. The Zero Trust  model earns trust instead of assuming it implicitly. It grants access to resources through authentication, authorization, and continuous monitoring, maintaining the necessary confidence in the legitimacy of users and devices before granting access. This security model minimizes the assumed trust and focuses on continuous and real-time validation and verification of entities seeking access to resources.

In the cybersecurity landscape, there are several influential ZT models. Well, BeyondCorp is a Google business security model for building zero trust networks that shift access controls to an individual device and user level. This allows workers to work from anywhere while knowing that the information is securely kept off the traditional VPN. Further, BeyondCorp allows access based on the credentials of users, be it their status or that of the devices, ensuring full authentication, authorization, and encryption independent of the network location. It provides fine-grained access to the resources of an enterprise, having multiple components that ensure whether a device is authenticated or not and a user authenticated therein to access desired applications \mbox{\cite{assunccao2019zero, alevizos2022blockchain}}. Microsoft moved to a zero trust access architecture, focused on least privilege principles among the identities, people, services, and IoT devices as it transitioned to cloud-based services and mobile computing. Thus, Zero Trust Access provides secure access by supplying comprehensive measures for device health monitoring, application and API controls, data encryption, and telemetry for threat and risk detection. The in-built next-generation segmentation, real-time threat protection, and analytics fortify the network security with a robust model for cybersecurity \cite{nair2023and,Anu:2021}. 
 The research framework of this study is predicated on the National Institute of Standards and Technology (NIST) Zero Trust (ZT) model. This choice is based on its accordance with the study's central focus on zero trust implementation within connected vehicle environments \cite{anderson2023zero,zhao2023research}. The NIST ZT model provides a robust foundation for addressing the complex security challenges inherent in vehicular networks.

\subsection{Zero Trust Principles}

The ZT model is founded on various core principles that significantly differ from traditional security models. The basic tenets of ZT security involve being proactive in network security by assuming that the network is under constant threat from hostile actors. This clearly indicates the need to recognize threats from both within and outside the network. It emphasizes that security risks can emanate from external sources such as hackers or malware, but they can also originate from compromised devices or malicious user intent. This clearly indicates that the mere location of network entities is not enough to base a trust assessment. Still, a multi-factor approach has to be taken and considered, which is above and beyond the physical or virtual location. Also, there is a need for strict authentication and authorization procedures for all devices, users, and network traffic to grant access to network resources. Besides, this model suggests that security policies should be dynamic. Instead of using static, predefined policies, this model relies on real-time data from multiple sources. This way, the network adapts to emerging threats by dynamically adjusting security controls in response to real-time evolving conditions \cite{he2022survey,gilman2017zero}.

Some related technologies used in the ZT model include Least Privilege Access\cite{haber2020zero}, which controls user access using just-in-time and just-enough access, adaptive risk policies, and data protection measures. Thus, security for both the data and productivity is enhanced. In addition, ZT networks incorporate micro-segmentation\cite{basta2022towards}, where security perimeters are broken into smaller zones, thus ensuring different access controls for different network segments. This ensures there is no lateral movement, and any compromised device or user account is quarantined once the presence of an attacker has been detected. ZT security emphasizes Multi-Factor Authentication (MFA), which requires users to present more than one form of authentication evidence to access resources, thereby enhancing authentication security beyond mere password entry. Encryption, a cornerstone of ZTA \cite{shore2021zero}, plays a crucial role in securing data at rest, in transit, or during processing. Encryption transforms sensitive data into a nonsensitive format, such as by replacing names with arbitrary identifiers, thereby minimizing the attack surface, significantly reducing the risk of data breaches, and providing effective protection.

\subsection{Zero Trust Architecture}
\begin{figure}[H]
  \centering
    \includegraphics[width=0.5\textwidth]{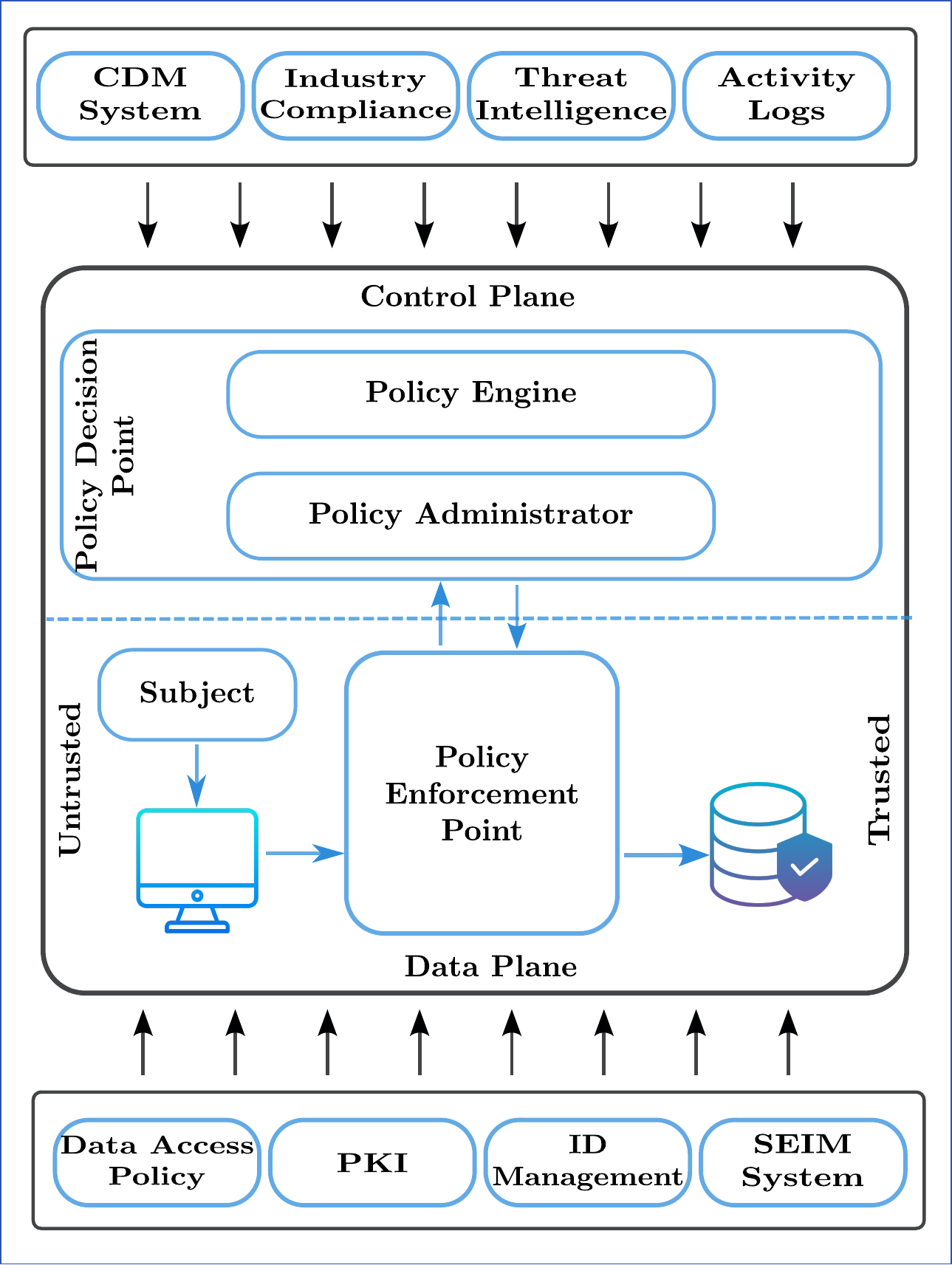}
    \caption{Logical components of the zero trust architecture (NIST SP 800-207 Model \cite{rose2020nist})}
    \label{fig:ZTA1-8}
    \end{figure}

Zero Trust Architecture is a cybersecurity model that implements ZT principles within a network of business infrastructure and operational procedures. It focuses on component interactions, workflow planning, and access controls to enhance security measures \cite{rose2020nist}. Figure \ref{fig:ZTA1-8} shows the main components of NIST's ZTA model. As depicted in this figure, the Policy Decision Point (PDP) is a crucial component for managing authentication decisions and access policies, consisting of two core elements: the Policy Engine (PE) and the Policy Administrator (PA). A PE is critical in making decisions about whether a certain subject should be allowed a resource by applying the Trust Algorithm (TA), the different external data sources, and business policies. It collaborates with a PA to develop and implement access policies. Both the PA and PE interlink with one another to allow or deny access based on PE decisions. To enforce policies, it might be integrated into the PE framework and communicate with the Policy Enforcement Point (PEP). The PEP is responsible for managing subject and enterprise resource connections, as well as communicating with the PA. It has the flexibility to function as a single component or be separated into client and resource sides, extending the trust zone beyond the PEP.\\

In a ZTA setup \cite{rose2020nist}, aside from the core components, various data sources, both internal and external, provide input and policy rules for the PE to use a TA in order to make decisions about access. The enterprise security infrastructure comprises several interconnected systems. The Continuous Diagnostics and Mitigation (CDM) system focuses on maintaining asset states, verifying factors like OS patching correctness, software integrity, and known vulnerabilities, and extending its reach to nonenterprise devices. The compliance system addresses regulatory frameworks and enforces policy rules to ensure industry compliance. Threat intelligence provides information on attacks, vulnerabilities, and malware to the policy engine. Network and system activity logs consolidate real-time insights into security status, while data access policies govern resource access. The Enterprise Public Key Infrastructure (PKI) generates and logs certificates that are potentially integrated with the global PKI ecosystem. The ID management system oversees user accounts, roles, and access details, possibly linked with PKI. Lastly, the Security Information and Event Management (SIEM) system collects security-centric data for analysis, refining policies, and issuing alerts for potential attacks on enterprise assets.

\begin{figure}[H]
  \centering
    \includegraphics[width=0.5\textwidth]{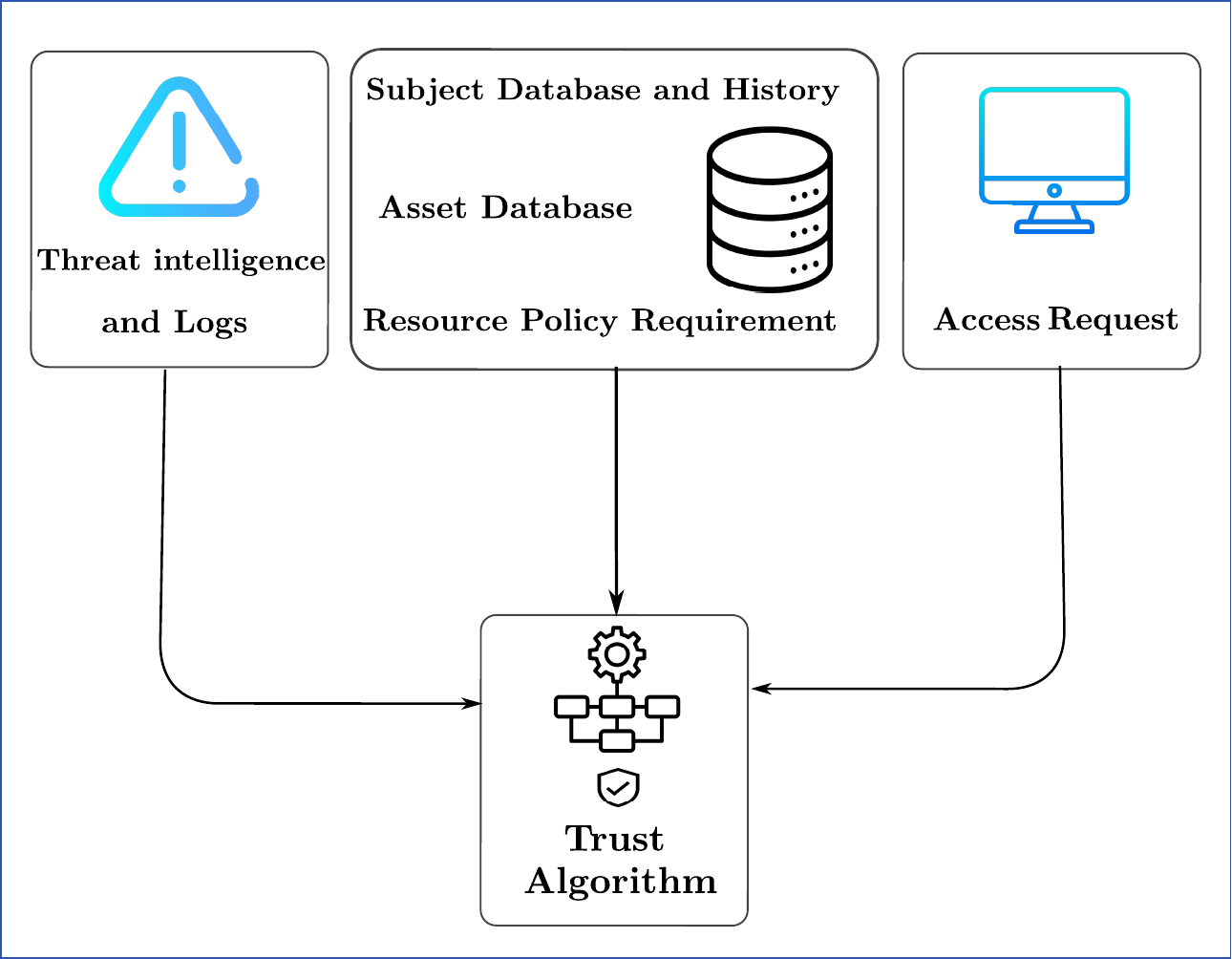}
    \caption{Trust algorithm \cite{rose2020nist}}
    \label{fig:trust3-9}
    \end{figure}

The policy engine in the ZTA is the central decision-maker, with its primary thought process guided by the TA. This algorithm, utilized by the PE, determines whether to grant or deny access to a resource. The decision-making process incorporates inputs from various sources, such as the policy database, which contains data on subjects, their attributes and roles, their past behavioral patterns, threat intelligence sources, and other metadata sources. This is illustrated in Figure \ref{fig:trust3-9}.
    
The access request process involves subjects requesting information about assets, considering factors like OS version, software, and security status. Access is restricted based on requester credentials and asset security posture. The subject database and history manage subject attributes and privileges, which are stored in the ID management system. The asset database tracks asset status, with access restrictions based on alignment with database information. Resource policy requirements establish minimum criteria for access, including authentication assurance levels and data sensitivity. Threat intelligence and logs provide information on threats and malware, typically managed by external services\cite{rose2020nist}.

\section{Zero Trust Model Implementation}
The implementation of the zero trust model generally involves deploying security measures and protocols that do not inherently trust any device or entity within a network \cite{itodo2024multivocal}, as described in the previous section, to provide robust defenses against evolving threats and vulnerabilities. The implementation process involves several stages and considerations, including the pillars that support the ZT model. In this section, we delve into the pillars of the zero trust model along with the Zero Trust Maturity Model (ZTMM) to provide clarity and guidance throughout the implementation process. Figure showcases the occurrence of zero trust implementation keywords

\begin{figure}
  \centering
    \includegraphics[width=0.7\textwidth]{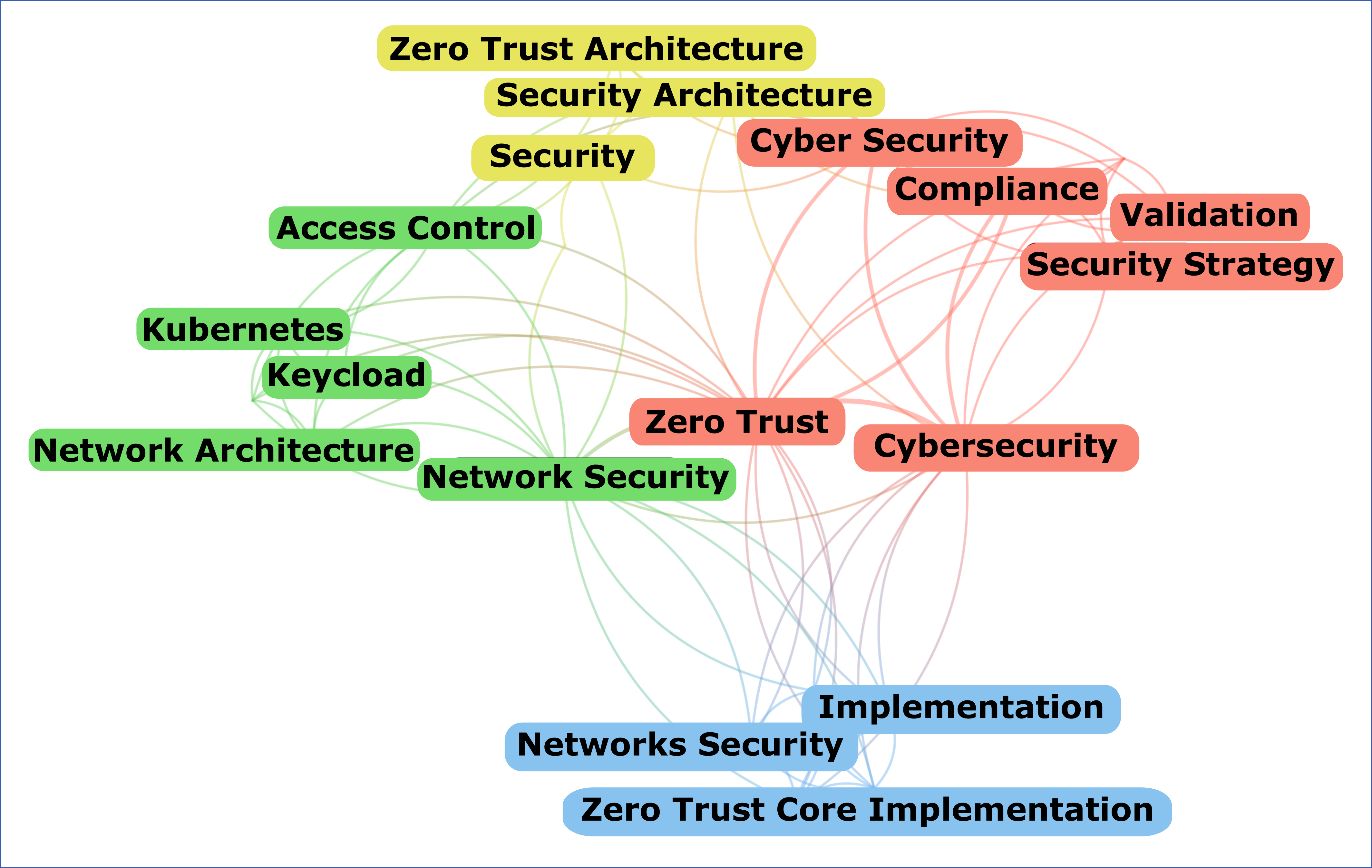}
    \caption{Zero trust implementation keywords occurrence}
    \label{fig:implementation}
    \end{figure} 

\newpage
\subsection{Pillars of Zero Trust Model}
\begin{figure}[H]
  \centering
    \includegraphics[width=0.5\textwidth]{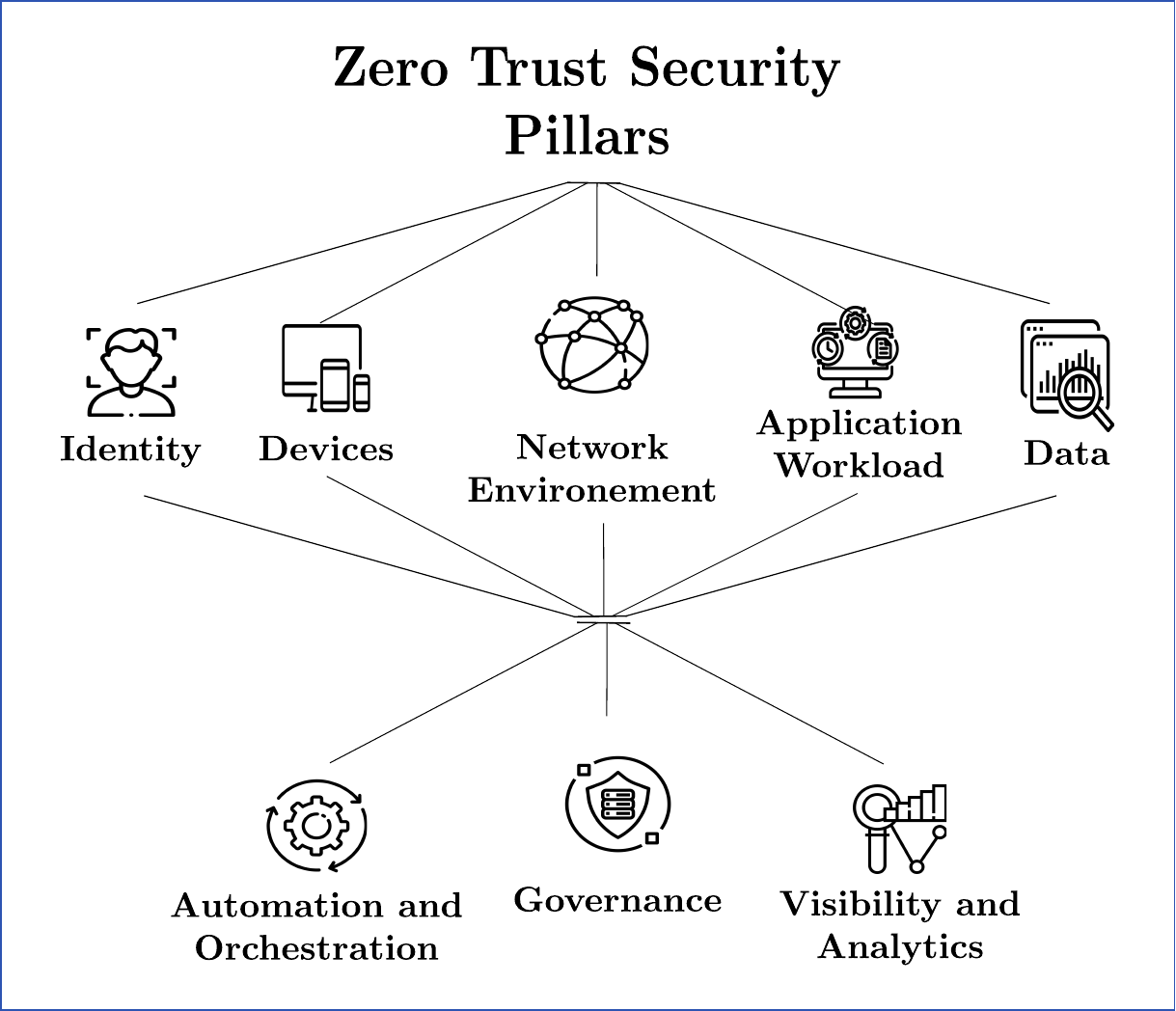}
    \caption{pillars of zero trust model \cite{CISA:2022:Online}}
    \label{fig:pillars010}
    \end{figure}

Before implementing ZTA, users must consider the seven main pillars that support a ZT model. The Cybersecurity and Infrastructure Security Agency (CISA)\cite{CISA:2022:Online} created a ZTMM (Figure \ref{fig:ztstages11}), which outlines the fundamental pillars essential for a robust ZT model. The pillars, as shown in Figure \ref{fig:pillars010}, were divided into Identity, Devices, Network/Environment, Applications/Workloads, And Data. Each pillar includes common components for visibility and analysis, automation and orchestration, and governance.

For which, identity describing the unique features of agency users or entities is essential to creating secure access controls. In addition, a number of diverse hardware assets, such as IoT devices and mobile phones, interconnect with the network, enabling seamless communication and exchanging data. Furthermore, applications and workloads that exist on-premises or in the cloud rely on this network infrastructure to function. Meanwhile, data, which is fundamental to agency operations, flows through these interrelated systems, placing a high value on data protection mechanisms. Moreover, visibility and analytics bring viewpoints to enterprise-wide events, guiding informed decisions and proactive security actions. Automation and orchestration smooth out security response functions; therefore, they enhance general efficiency and resilience. Last but not least, governance ensures that enforcement of cybersecurity policies runs across the enterprise, further strengthening adherence to ZT principles and federal requirements.

\subsection{Zero Trust Maturity Model}
That would be a progressive ZT implementation, a procedure requiring time and effort, so it could not be implemented quickly. Most networks can use the majority of their installed base by implementing ZT concepts, but achieving a mature ZTA often requires additional capabilities to fulfill its promise completely. It is not imperative to transition to a mature ZTA in one stroke. Instead, integrating ZT functionality incrementally as part of a strategic plan helps reduce risks at every stage. As the ZT implementation evolves, improved visibility and automated responses empower defenders to counter emerging threats effectively. The National Institute of Standards and Technology (NIST)\cite{rose2020nist} has published a ZTA guide outlining the seven tenets of the ZTMM. These tenets are described as follows:

\begin{enumerate}
\item  Considering as resources all data sources and computing services.
\item Securing every communication across network locations.
\item Per-session access granting for individual enterprise resources.
\item Resource access governed by dynamic policies.
\item Integrity and security posture of the enterprise assets are monitored and assessed comprehensively.
\item Strict enforcement of dynamic authentication and authorization for resource access.
\item Enhancing security posture through comprehensive data collection on assets, infrastructure, and communications.
  \end{enumerate}

  \begin{figure}[H]
  \centering
    \includegraphics[width=0.5\textwidth]{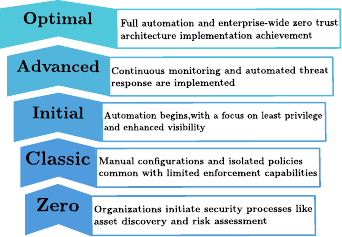} 
    \caption{Stages of zero trust maturity model \cite{CISA:2022:Online}}
    \label{fig:ztstages11}
    \end{figure} 
    
The NIST outlines that the complete and pure implementation of all tenets may not be entirely achievable within a given strategy. CISA's ZTMM is just one of the diverse options available to facilitate the shift towards a ZT model. This model is based on the pillars discussed earlier (subsection 3.1).

The zero trust journey unfolds in five stages (Figure \ref{fig:ztstages11}). In stage zero, organizations initiate security processes like asset discovery and risk assessment. Transitioning to stage classic, they often rely on manual configurations and isolated policies with limited enforcement capabilities. Progressing to stage initial, automation begins, emphasizing least privilege and improved visibility. Advanced stages prioritize continuous monitoring and automated threat response, ensuring centralized visibility and integrated policies. Finally, stage optimal marks the pinnacle, where full automation and enterprise-wide implementation of ZTA are achieved, demanding comprehensive revisions across people, processes, and technology to complete the initiative.

\subsection {Open-source Tools For Zero Trust Model}
Various organizations have developed a wide array of open-source tools to facilitate the implementation of zero-trust security architectures. These tools encompass a range of functionalities crucial to the Zero Trust model, including micro-segmentation, fine-grained dynamic access control, safe communication, and centralized policy management. These features collectively simplify the application and enforcement of Zero Trust principles across diverse IT environments.
\begin{itemize}
  
    \item {OpenZiti}\\
   OpenZiti is an Open Source project by NetFoundry that allows ZT integration into existing systems using a very flexible ZT overlay network. This gives a private networking capability that is unavailable to the Internet and makes port sniffing ineffective in finding any unauthorized access owing to the fact that there are no listening ports on the services. Strict identity-based access control insists on strong authentication before establishing a connection. It also supports Zero Trust methodologies such as attribute-based access control (ABAC), end-to-end encryption, and least privilege \cite{kanerva2024lateral}.
   
   \item {Keycloak}\\
   Keycloack is an open-source Identity and Access Management system used in securing web applications, microservices, and APIs. It provides a single location where authentication and authorization are managed and has a large community of users. Keycloak is considered to be one of the top IAM solutions within the open-source space and is very fast on its way to gaining high popularity due to its powerful capabilities it has and speed of development \cite{bartovsadaptive}. Noticeably, this aligns with Zero Trust, a model in which IAM plays a very active role since it assures and authenticates secure access to networks and applications.

      \item {Tailscale}\\
      It is a modern implementation of Zero Trust VPN, creating secure tunnels to services and applications across untrusted networks. Each host has to run a client software while having some central server coordinate authentication keys and the access policies. Clients form a secure mesh network. Concretely, Tailscale creates full VPN tunnels for all client traffic and is based on the Wireguard protocol\cite{hilbig2023state}.

      \item {Ockam}\\
      Ockam is a library that enables secure end-to-end communication. It works as a Policy Enforcement Point (PEP) to build Zero Trust applications in Rust and supports key establishment, rotation, revocation, and attribute-based access control mechanisms. Ockam works with many transport protocols like TCP, UDP, WebSockets, and Bluetooth. All Ockam applications create secure channels using either the Noise Protocol Framework or the X3DH protocol\cite{hilbig2023state}. 
\end{itemize}

\section{Zero Trust Applications}
Zero Trust principles, characterized by their proactive and continuous verification model to security, have found widespread application across diverse fields, such as cloud computing, IoT, 5G/6G, and connected vehicles, each presenting its own challenges and requirements. 

In cloud computing, ZT principles play a crucial role in ensuring continuous verification of access to cloud resources, thereby minimizing the risk of unauthorized access and data breaches. Various studies have contributed to advancing the understanding and implementation of ZT model in different contexts. \textit{Ahmed et al.} \cite{ahmed2020zero} proposed a decentralized identity and access management (IAM) framework to minimize the controlling power of any single entity over digital assets in cloud environments. Likewise, \textit{Chuan et al.} \cite{chuan2020implementation} introduced a technique that helps small and medium-sized businesses establish situations where users access intranet data across servers hosted in the cloud. \textit{Mehraj et al.}\cite{mehraj2020establishing} created a conceptual model to address security issues that appear while deploying new infrastructure in the cloud. This model allows cloud server providers and customers to identify legitimate entities inside the cloud environment. Leveraging machine learning for behavior analysis. \textit{Saleem et al.} \cite{saleem2023secure} propose a zero trust security architecture that uses the Rich model for trust verification in SaaS. Furthermore, \textit{Ferretti et al.} \cite{ferretti2021survivable} addressed the issues around the zero trust principles and trusting control plane components by proposing a method to divide trust and mitigate potential attacks. For wireless environments, \textit{Mandal et al.} \cite{mandal2021cloud} presented a new access control model based on a zero trust network. This model aims to reduce vulnerabilities, focusing on MAC spoofing prevention and general security reinforcement. Additionally, in order to strengthen network security and protect cloud assets, researchers in \cite{decusatis2016implementing} suggested a network design that included first-packet authentication and transport access control. \textit{Rodigari et al.} \cite{rodigari2021performance} evaluated the performance of ZT implementation with a focus on multi-cloud scenarios using Istio service mesh, providing an understanding of resource use and latency. Finally, \textit{Sarkar et al.} \cite{sarkar2022security} presented a combined zero trust network architecture for cloud networks that coordinates user security, improves network visibility, slows down cyberattacks, and automates trust computations.\\

IoT security depends on ZT principles, allowing stringent access controls and device authentication to reduce risks. \textit{Awan et al.} \cite{awan2023blockchain} suggested a dynamic method that combines blockchain technology, attribute-based access control (ABAC), and ZT principles to improve security in smart cities and other large-scale IoT infrastructures. Their solution ensures all security features, including data privacy, confidentiality, and authentication, for every user accessing IoT devices across a network. \textit{Dimitrakos et al.} \cite{dimitrakos2020trust} presented a unique ZTA-based trust-aware continuous authorization technique created specifically for consumer IoT networks, especially smart home situations. This lightweight architecture provides effective speed, scalability, and flexibility while answering privacy and security concerns. In another study, \textit{Shah et al.} \cite{shah2021lcda} proposed a lightweight continuous device-to-device authentication protocol (LCDA) that will be suitable for low-resource IoT devices. Through process simplification and lightweight encryption, the protocol reduces energy consumption while ensuring strong resistance against a wide variety of threats. For IoT contexts,\textit{Wang et al.} \cite{wang2023secure} developed a zero trust architecture-based safe access technique that exploits identity-centric authentication and dynamic access control. Their model detects and mitigates cyber risks in distributed power access settings, while balancing comprehensive security measures and lightweight architecture. To address the complexity of IoT access control, \textit{Meng et al.} \cite{meng2022continuous} developed a blockchain-based continuous authentication system inside a ZT model for solving the complexity of IoT access management. That system improves efficiency by ensuring safety and categorizing the devices into three categories: trustworthy, suspicious, and untrusted. To deal with the exponential increase in data within 6G networks, \textit{Han et al.} \cite{han2022zt} developed a blockchain-based zero trust data storage technique. The proposed method reduces resource consumption, improves data security, and increases storage and bandwidth. Moreover, \textit{Zhao et al.} \cite{zhao2021blockchain} proposed the IoT device authentication system based on blockchain, which brings not only security but categorizes devices as entities of trust and provides strong options for authentication.

Similarly, using ZTA in 5G and 6G networks tackles comparable security concerns for the upcoming generation of fast, linked networks. In this context, ZT provides robust security measures to safeguard critical communication channels against emerging threats. Coping with the escalating connectivity of devices in untrustworthy environments within 5G/6G networks necessitates dynamic security frameworks. The authors in \cite{ramezanpour2022intelligent} proposed an intelligent Zero Trust Architecture (i-ZTA) for 5G/6G networks. The i-ZTA provides dynamic network assurance, real-time processing of big data, AI-driven components for ZT principles, and presents innovative research directions for enhancing information security in untrusted 5G/6G networks.
A blockchain-based zero trust security-enabled federated learning system, "Shunk", was suggested by \cite{bandara2022skunk} to deal with the challenges of configuration complexities and security management in 5G/6G networks with network slicing and the limitations of existing centralized federated learning (FL) systems. The platform ensures transparency and data provenance. Its sharding-based architecture improves security across various network slices, addressing centralized coordinator-based federated learning challenges. Scientists in \cite{sedjelmaci2023zero} recommended a robust ZTA empowered intrusion detection model to resolve challenges in 6G edge computing. The model ensures accurate cyber-attack detection, protection from internal and external threats, and efficient network cost management, as demonstrated by high detection rates, low false positive rates, and reduced network costs in simulation results. Scholars in \cite{sedjelmaci2023distributed} proposed an Intelligent Zero Trust (iZT) model for enhanced security in 6G networks that employs distributed threat detection and collaboration to bolster defences against RAN threats.
\textit{Hireche et al.}\cite{hireche2022deep} presented a distributed trustworthy SelfDN utilizing zero trust telemetry, automated code rewriting with P4 and AI, and secure, decentralized knowledge sharing with blockchain to counter challenges in complex networks in the face of high connectivity. The model enhances monitoring with real-time telemetry, achieves autonomous code translation, enables decentralized knowledge sharing, demonstrates practical effectiveness against distributed denial of service (DDoS) attacks, and positions AI as a transformative element in achieving the SelfDN vision.\\

\begin{figure}
  \centering
    \includegraphics[width=0.7\textwidth]{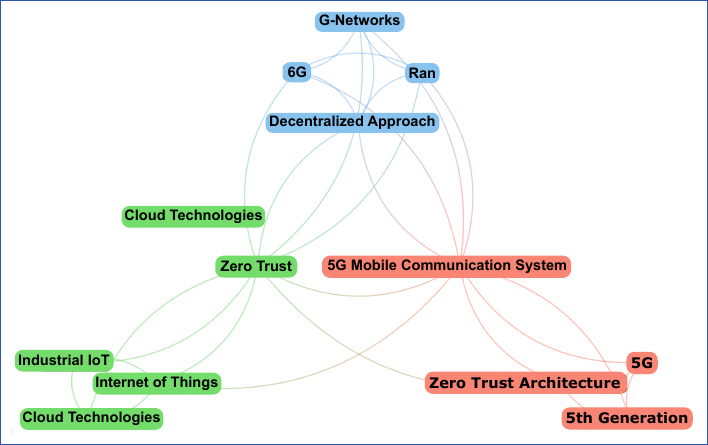}
    \caption{Zero trust applications keywords occurrence}
    \label{fig:ztstages12}
    \end{figure}

Transitioning from the application of ZT principles in cloud computing, IoT security, and 5G/6G networks to their integration in connected vehicles highlights the broad spectrum of cybersecurity threats faced by these emerging technologies. Implementing trust management in vehicular environments poses challenges due to frequent network disconnections, dynamic mobility patterns, and constraints in computing and communication capabilities \cite{el2019trust}. Nonetheless, ZT effectively addresses these risks by enforcing authentication mechanisms and segmenting network access, providing a proactive model to improving network security in connected vehicles. \textit{Zayed et al.} \cite{zayed2022owner} proposed implementing a zero trust architecture through verifying vehicle owner identity.
Researchers in \cite{li2023zero} presented a security scheme founded on blockchain and zero trust principles to protect both the chargers and the cloud platform. On the other hand, \textit {Zhao et al.} \cite{zhao2023research} suggested a security model for smart charging stations, using a ZT model to secure vehicle-node communication. \textit{Wang et al.} \cite{wang2023attribute} introduced a Zero Trust Access Control Model (AU-ZTAC) based on attributes and user trust scores. Furthermore, scientists in \cite{fang2022zero} introduced a security scheme for the Internet of Vehicles (IoV) that combines multifactor authentication, blockchain encryption, and a zero trust architecture. \\
Figure \ref{fig:ztstages12} demonstrates the occurrence of the zero trust application keywords across the different fields explained earlier.

\section{Cybersecurity Challenges in Connected Vehicles}
In this section, we delve into the critical aspects of securing CVs, beginning with an overview of the topic. Our investigation encompasses a range of considerations, including trust models and intra-vehicle and inter-vehicle network cyber-attacks. First, we introduce the overview of connected vehicles to establish the necessary context for understanding the complexities and variations involved in ensuring the security of these intelligent transportation systems.

\subsection{Connected Vehicles Overview}

Connected vehicles are vehicles that use communication technologies to enable them to exchange data with the Internet, other vehicles, infrastructure, and mobile devices wirelessly. They provide features such as real-time data transmission, remote control, and applications like traffic safety, infotainment, and autonomous capabilities. These vehicles enhance situational awareness for occupants, playing a crucial role in ITS and IoV. Vehicle networks, both external and internal, facilitate communication between different systems within CVs, in both intra and inter-vehicle communication, ensuring seamless connectivity and collaborative functionality.

\subsubsection{Intra-vehicle communication}

Intra-vehicle communication entails data and information sharing across various vehicle devices, encompassing control units, monitoring components, mobile devices, execution components, wearables, and other elements.
A complex diverse network is formed by the intra-vehicle network, enabling information exchange and coordinated control among electronic components, including sensors (LiDAR, RADAR, Camera, GPS...etc), actuators, and Electronic Control Units (ECUs) for an enhanced driving experience\cite{chen2023towards}.
The intra-vehicle network comprises interconnected subnets communicating through gateways with various protocols. The proliferation of electronic components, particularly numerous ECUs in modern vehicles connected to the intra-vehicle network (IVN), is vital for advancing autonomous vehicle technology, yet it raises safety concerns with an expanded attack surface \cite{karopoulos2022demystifying}.\\
The current automotive landscape incorporates five key intra-vehicle networking technologies: Controller Area Network (CAN), Local Interconnect Network (LIN), FlexRay, Media Oriented Systems Transport (MOST), and Automotive Ethernet. \\

The Controller Area Network (CAN) is a cost-effective and reliable serial bus network connecting devices within control applications. It supports data rates up to 1 Mb/s, utilizing a message-based broadcast protocol for intra-vehicle communication among ECUs. However, limitations include relatively low bandwidth and shared data transmission, restricting its application in certain domains \cite{bozdal2020evaluation, song2020vehicle}. The Local Interconnect Network (LIN), standardized by ISO 17987, is a cost-effective network for low-bitrate vehicular applications, operating at 1 to 20 kbps. It utilizes a single wire to connect a master to multiple slaves, serving as an affordable solution for automotive sensor and actuator connections, linking them with higher-level networks like CAN \cite{aliwa2021cyberattacks}. FlexRay provides higher bandwidth and fault tolerance than CAN, albeit at a higher cost. Its dual-channel architecture makes it suitable for safety-critical systems like brake-by-wire, offering a competitive alternative for high-performance applications \cite{mahmood2020connected}. Media Oriented Systems Transport (MOST) excels in transmitting multimedia data within vehicles, with bandwidths ranging from 25 to 150 Mbps. MOST operate in a ring topology, supporting up to 64 devices, but its higher cost limits its use to applications requiring essential multimedia capabilities \cite{rajapaksha2023ai}. As automotive ethernet evolves to meet diverse needs like bandwidth and network management, it provides enhanced communication bandwidth for advanced driving features and infotainment systems. Despite its standardization, automotive ethernet inherits security weaknesses from traditional ethernet, necessitating robust cybersecurity measures to safeguard intra-vehicle networks. While each protocol offers advantages, its vulnerability to cyber attacks underscores the need for stringent cybersecurity to ensure modern automotive safety and the integrity of the system \cite{tiberti2023hybrid, lee2020design}.

\subsubsection{Inter-vehicle communication}
Inter-vehicle communication (IVC) involves real-time information sharing among vehicles within a network facilitated by modern technologies. In the context of ITS and autonomous driving, this communication kind crucial for sharing safety-related data and enhancing perception and planning. It allows vehicles to autonomously respond to potential risks, contributing to accident prevention without direct driver intervention. Given traditional limitations of sensors and reliance of autonomous vehicles on environmental data, secure and reliable communication technologies are essential for effective inter-vehicle networking \cite{wang2018networking,mejri2014survey}. 
VANET extends the principles of traditional MANET (Mobile Ad-hoc Network) to traffic scenarios, which aim to swiftly transmit safety messages and infotainment services for drivers and autonomous vehicles \cite{sharma2019survey, lee2021vanet}. Prioritizing critical safety information delivery, VANETs offer various services, including safety applications, driving assistance, cooperative driving, collision avoidance, traffic control, and entertainment \cite{hussein2022comprehensive, luckshetty2016survey}. It utilizes V2X (Vehicle-to-Everything) technologies, including V2V (Vehicle-to-Vehicle), V2I (Vehicle-to-Infrastructure), I2I (Infrastructure-to-Infrastructure), V2P (Vehicle-to-Pedestrian), and V2C (Vehicle-to-Cloud).

In a VANET setup, vehicles utilize onboard units (OBUs), roadside units (RSUs), and a trusted authority (TA). The TA, an external device, registers RSUs, authenticates vehicles, and monitors the network via RSUs \cite{khelifi2019named,arif2019survey}. RSUs, stationed along the roadside and utilizing wireless technologies like WiFi, act as intermediaries between the TA and vehicle OBUs, conveying safety instructions to nearby vehicles. OBUs, fixed on vehicles, facilitate the transmission of vital messages to neighboring vehicles \cite{mundhe2021comprehensive}.\\

The increase in vehicle traffic highlights the significance of VANETs within ITS. Although VANETs facilitate wireless communication among vehicles, their constrained processing capabilities confine them to short-term or localized applications, hindering the advancement of vehicle-based services. To address this limitation, an integrated network infrastructure is required. The incorporation of IoT is reshaping conventional VANETs into the IoV, distinguished by seamless wireless connectivity among vehicles \cite{assem2023data}. This evolution promises to enhance road safety and reduce congestion while improving the driving experience.

V2V communication via DSRC or IEEE 802.11p creates a local ad-hoc network with limited reach and capacity, supporting around one hundred vehicles within a few hundred meters. However, 5G's Cellular-Vehicle to Everything (C-V2X) technology can also establish V2V connections. When comparing V2V using IEEE 802.11p to C-V2X, it's evident that C-V2X offers a broader range. However, there could be a trade-off regarding data capacity \cite{bazzi2017performance, mannoni2019comparison}, as the Radio Access Network (RAN) coordinates participants in the network \cite{arthurs2021taxonomy}. Despite the intended objectives of 5G technology to achieve great capacity, massive bandwidth, vast connections, low latency, and low power consumption, challenges arise due to vendor-specific hardware and protocols. As 5G networks become more sophisticated, operators face difficulties dynamically adjusting network operations. This complexity and the associated costs become more pronounced as the technology evolves \cite{mahi2022review}. 
On the other hand, integrating 6G technologies into ITS is poised to revolutionize network performance. This includes advanced communication technologies like AI-enabled wireless networks, ultra-reliable low-latency communication, solutions for faster mobility, and secure practices for IoV. 6G integration promises to improve efficiency, smart traffic management, and address dynamic challenges in transportation networks, presenting opportunities for advanced connectivity and streamlined data processing \cite{adhikari2021roadmap}. Furthermore, the onboard server has limited computing and storage capability, and intricate computing tasks are expected to be accomplished with the support of cloud computing or mobile edge computing\cite{guo2024survey}.

Vehicular clouds play a crucial role in providing services such as route planning and safety control to vehicles. These clouds communicate with other commercial clouds as well as sensor clouds, forming based on the services needed. Core functionalities, including communication, processing, sensing, and storage, are essential for vehicular clouds. Vehicles can join existing clouds or initiate formation, and a dynamic cloud leader manages resources and incoming requests. This technological model offers feasible solutions for enhancing vehicular cloud infrastructure\cite{ahmad2017characterizing}.

\begin{figure}[H]
  \centering
    \includegraphics[width=0.6\textwidth]{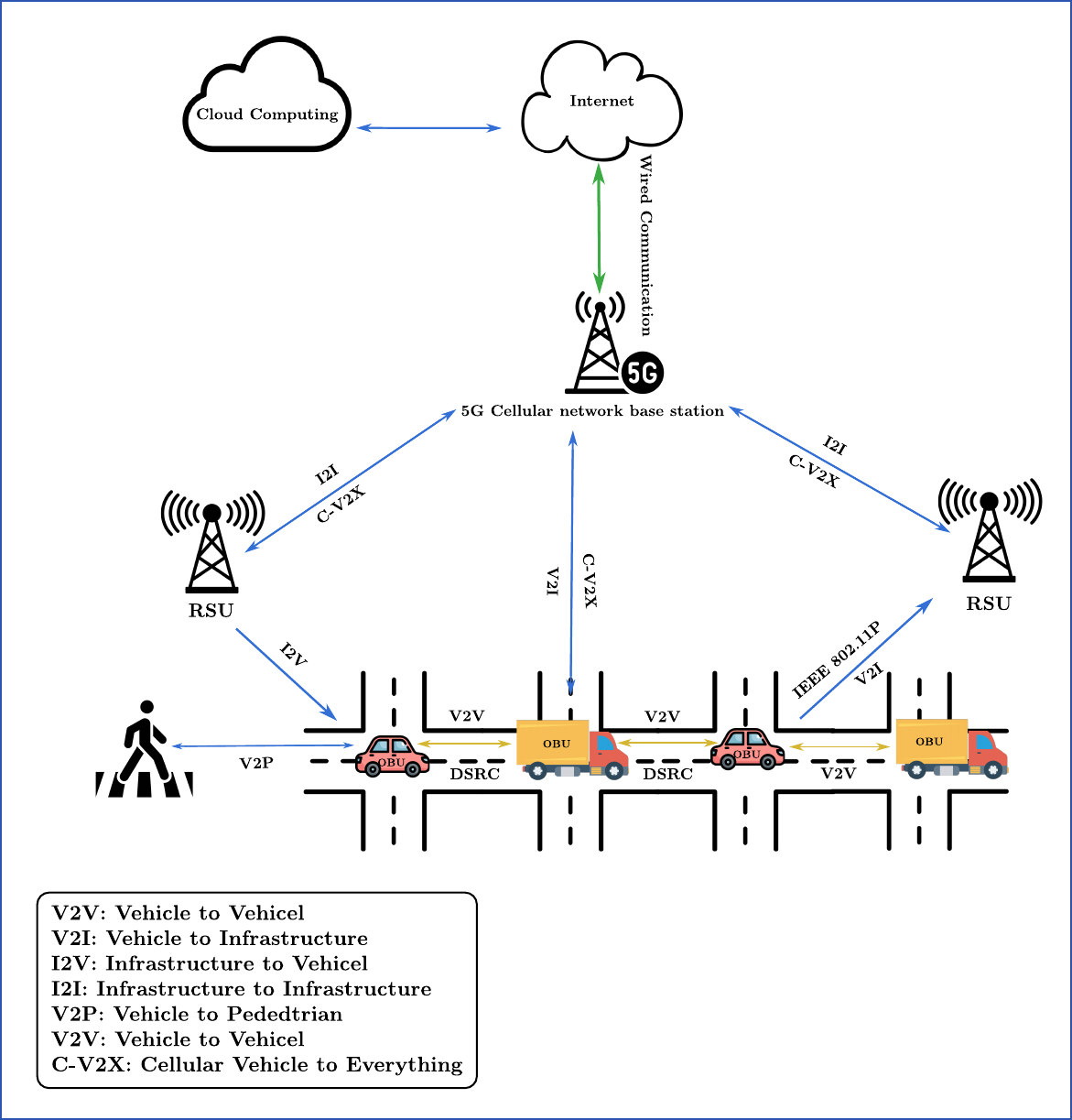}
    \caption{Hybrid VANET architecture (cloud computing, 5G-cellular networks)}
    \label{fig:vanet architecture13}
    \end{figure}

Figure \ref{fig:vanet architecture13} represents A VANET architecture with cloud computing and 5G C-V2X in the Internet of Vehicles environment. VANET architectures typically consist of interconnected elements to facilitate efficient communication and data processing in a vehicular network.

\label{sect: VANET ARCHITECTURE}

\subsection{Intra-vehicle Networks Cyber-attacks}
Intra-vehicle networks (IVNs) face a range of cybersecurity threats. Cyber attackers can exploit vulnerabilities by obtaining physical access through entry points like the OBD-II port, USB connections, and CD player. Moreover, short-range wireless technologies such as Bluetooth and RFID and long-range options like Wi-Fi and LTE offer potential avenues for intrusion. Additionally, attackers may focus on exploiting vulnerabilities in communication protocols and the sensors, including LiDAR, RADAR, and cameras, integrated into IVNs. In this subsection, we briefly explore the main IVN cyber attacks including sensors and protocols cyber attacks.

\subsubsection{Intra-vehicle sensors cyber attack} 
 Cyber threats to connected vehicles are a significant concern, particularly for key components of the vehicle. Research in \cite{saeed2023review}  describes the various attacks on intra-vehicle sensors, like the camera \cite{el2020cybersecurity}, 
integral for functions like lane detection and pedestrian recognition. The camera in CVs faces susceptibility to manipulation by attackers. Deceptive tactics involve introducing ambiguous shapes on road signs or disrupting lane detection using unconventional elements. One specific threat, a blinding attack, uses a powerful laser beam to increase tonal values and hide the camera feed, causing complete blindness to vehicle sensor inputs. This can cause vehicle deformation and need emergency braking. Another form of attack, the auto-control attack, focuses on camera sensors, employing continuous bursts of light to manipulate auto controls and prevent image stabilization. Typically executed as a front, rear, or side attack, this assault can hinder the functionality of the camera.

LiDAR, crucial for precise object detection in safe transportation, is vulnerable to cyber-attacks, especially in its fixed position within Adaptive Cruise Control (ACC) and collision avoidance systems. Despite the limitations of the camera, the high spatial resolution of LiDAR allows accurate scanning to distinguish between cars and pedestrians. However, cyber threats pose risks to the reliability of LiDAR.
Denial of service attacks can slow down object detection by flooding the system with excessive objects. Moreover, replay attacks involve recording LiDAR signals for later transmission, thereby creating non-existent objects that disrupt the accuracy of distance gauging in LiDAR. Furthermore, blinding attacks saturate LiDAR with light, discreetly denying its services. Additionally, spoofing attacks make LiDAR detect non-existent objects, potentially causing miscalculations. Lastly, jamming attacks disrupt the operation of LiDAR by emitting light at the scanner unit, further complicating its functionality \cite{saeed2023review, el2020cybersecurity}.

Cyber attackers exploit vulnerabilities in the Global Positioning System (GPS) \cite{saeed2023review, el2020cybersecurity} through spoofing attacks, manipulating GPS data to misdirect vehicles. The susceptibility of Tesla vehicles to wireless GPS spoofing underscores the severity of this security threat. Spoofing involves strengthening false signals over authentic GPS signals, leading vehicles off course. Simpler attacks like jamming, where noise disrupts authentic signal reception, also pose risks to vehicle navigation systems. In GPS jamming attacks, signals are intentionally disrupted, preventing the vehicle from being located. Spoofing attacks take advantage of the low power of GPS signals, supplying false signals to compromise data integrity. Such attacks can be a preliminary step for additional threats, including replay and tunnel attacks. 

Radar sensors \cite{el2020cybersecurity} serve diverse functions: long-range for ACC, mid-range for Lane Change Assistants (LCA), and short-range for parking obstacle alerts. However, security concerns arise, and jamming attacks involve disrupting radar sensors with a signal in the same frequency band, reducing the signal-to-noise ratio of the sensor and impairing its capacity to identify objects in nearby areas. Risks also occur due to spoofing relay attacks, in which malicious actors continuously retransmit a previously legitimate signal in order to falsify it. Attackers store and duplicate a signal using a Digital Radio Frequency Memory (DRFM) repeater to trick the radar into thinking it is legitimate. The integrity of radar-based systems is being challenged by these attacks.

\subsubsection{Intra-vehicle communication protocols cyber attacks}

The integration of various components within the intra-vehicle network has enhanced safety, comfort, and performance while connecting with external devices, which further enriches the user experience. However, these digital communications also present security risks \cite{anwar2023security}. This section illustrates the attacks that compromise the CAN, LIN, MOST, FlexRay, and automotive ethernet networks.

Conducting a vulnerability assessment of a network is crucial for identifying security issues. For the CAN protocol, such an assessment focuses on confidentiality, integrity, and availability. Unfortunately, the protocol lacks inherent cryptographic methods, compromising confidentiality and allowing unauthorized access to sensitive data. While the CAN bus employs  Cyclic Redundancy Checksum (CRC) for integrity verification, it is unable to prevent malicious data injection, thus failing to ensure data integrity. In addition, the priority-based messaging system threatens availability and could result in the inability of the network to be available to other lower priority nodes. Therefore, the CAN protocol performs poorly on all three security criteria and thus is vulnerable to attacks from various attacks \cite{bozdal2020evaluation}.

The CAN bus security system is considered sensitive because of the variety of potential attacks, including bus-off attacks, DoS, masquerading, injection, eavesdropping, and replay attacks. The CAN frames lack encryption and authentication mechanisms, which is why these attackers take advantage of them by disrupting communication and also compromising critical systems in a vehicle \cite{zhang2023bit}. Through their error message flooding, the bus-off attacks may cause nodes to go into a "bus-off" state where the communication between critical vehicle systems shuts down. DoS attacks would block all communication over the bus, violating the CAN bus's availability. In contrast, man-in-the-middle attacks violate the integrity of communication by sniffing and modifying the messages. Injection attacks, conversely, through the exploitation of the OBD-II ports, gain unauthorized access to the ECUs. In contrast, replay attacks disrupt vehicle operations by causing the replaying of valid frames in repetition, this violates the freshness of the intra-vehicle network property. Masquerading attacks exploit unencrypted communication patterns of CAN frames, and eavesdropping compromises privacy, allowing unauthorized access to intra-vehicle information and facilitating subsequent attacks such as masquerading, injection, and replay. Implementing robust countermeasures, such as encryption and authentication mechanisms, is essential for enhancing cybersecurity resilience and defending against these multifaceted attacks in the CAN bus system \cite{anwar2023security,rathore2022vehicle}.

LIN bus attacks can impact non-critical vehicle systems, affecting functionalities like infotainment, climate control, or lighting. 
Response collision exploits vulnerabilities in the physical and data link layers of the LIN, causing message collisions and potential disruptions. Header collision manipulates header fields, compromising the arbitration process and leading to conflicts in simultaneous transmissions \cite{takahashi2017automotive}. Injection attacks exploit application layer vulnerabilities, allowing malicious data injection and system malfunctions. Forced Suspend Mode attacks temporarily suspend targeted nodes, disrupting communication and leaving them vulnerable. DoS attacks take advantage of the lack of authentication by continually performing header collisions, causing the bus to shut down and denying service. These attacks highlight the need for strong security measures in LIN systems, especially when critical safety functions are compromised \cite{anwar2023security} . Finally, The purpose of a message spoofing attack is to disrupt vehicular communication by sending unauthorized, false messages and exploiting vulnerabilities in the master-slave model of the LIN \cite{rathore2022vehicle}. 

Attacks on the MOST bus, integral for high-end multimedia functions in vehicles, can severely impact multimedia and infotainment systems, leading to compromised audio and video playback, connectivity loss, navigation malfunctions, and disruptions in entertainment features. Synchronization disruption, injection, denial of service, and jamming attacks pose significant threats. Synchronization disruption involves introducing false timing information to disrupt MOST synchronization. Injection attacks aim to inject malicious data or commands, thereby altering normal functioning. DoS attacks target the data link layer, delivering false messages to interrupt legitimate ones, and jamming attacks disrupt low-priority legitimate messages \cite{anwar2023security, rathore2022vehicle}.

FlexRay bus attacks pose serious risks to safety-critical vehicle systems, impacting real-time communication for functions like braking and steering. Potential outcomes include compromised control signals, disrupted ECU communication, and system failures. Common FlexRay attacks include masquerading, eavesdropping, injection, replay, man-in-the-middle, full DoS, and targeted DoS.
Masquerading involves unauthorized devices being presented as legitimate ones and gaining access to sensitive data. Eavesdropping sees attackers intercept communication without consent, potentially accessing sensitive information. It affects data confidentiality and includes security primitives. Injection attacks manipulate frames between network nodes, causing system malfunctions. Replay attacks intercept and retransmit valid messages, misleading the receiver \cite{kim2020vehicle}. Man-in-the-middle attacks occur when an attacker manipulates communication between nodes, leading to security breaches. Full denial of service overwhelms the FlexRay network with dominant signals, disrupting legitimate messages. By transmitting dominant signals, targeted denial of service prevents specific, legitimate messages. These attacks exploit vulnerabilities like a lack of encryption and authentication 

Automotive ethernet networks are increasingly integral for high-bandwidth applications like infotainment, advanced driver-assistance systems, and autonomous driving. However, potential attacks on these networks pose serious risks, including unauthorized access, critical command manipulation, and communication disruption among ECUs. Consequences range from compromised vehicle functionality to safety risks and privacy breaches.

Network access, eavesdropping, spoofing, injection, replay, Address Resolution Protocol (ARP) cache poisoning, DoS, TCP hijacking, and CAM table overflow are common attacks on automotive Ethernet. Unauthorized access is a component of network access attacks.
Efforts at access, endangering control systems and data privacy. By intercepting unapproved network access, eavesdropping jeopardizes privacy. Injection attacks happen when an attacker inserts malicious messages into the automotive ethernet, compromising the system; replay attacks take advantage of the lack of authentication by intercepting and retransmitting legitimate packets to cause malfunctions. Spoofing entails impersonating a device to obtain unauthorized access because there is no authentication. False ARP messages are the cause of ARP Cache Poisoning, which can result in Man-in-the-Middle attacks. DoS attacks overwhelm the network, disrupting the system. TCP hijacking involves intercepting and controlling TCP connections. CAM table overflow attacks exploit limited table size, potentially causing system crashes or DoS\cite{anwar2023security}.

\subsection{Inter-vehicle Networks Cyber-attacks}
Inter-vehicle communication in modern connected vehicles enables data exchange, supporting safety features like collision avoidance and cooperative cruise control. However, this connectivity also introduces cybersecurity risks, making vehicles potential cyberattack targets. Malicious actors could exploit vulnerabilities in communication protocols, compromising data integrity and posing security threats.
In the upcoming section, we will explore various attacks on VANET networks, which aim to compromise availability, confidentiality, integrity, authenticity, and nonrepudiation within vehicular networks. 

Ensuring information availability is critical in VANETs, where various threats and attacks can impact this aspect. DoS attacks, executed by internal or external malicious nodes, disrupt communication channels and render services unavailable, posing a significant threat to VANET applications \cite{mejri2014survey}. Jamming Attacks involve using a powerful signal to block valid safety alerts, posing a serious threat to safety applications in VANETs \cite{arif2019survey}. Blackhole attacks, present in both VANETs and ad hoc networks, involve malicious nodes throwing away packets of data to cause a DoS. Malware attacks infiltrate VANETs through software components responsible for operating OBUs and RSUs. Broadcast tampering attacks involve transmitting fake messages to authorized vehicles, hiding correct safety messages, and posing a risk of dangerous accidents. Greedy behavior in VANETs involves deceiving others with false data, causing bandwidth issues and delays for registered users. Spamming attacks flood the network with numerous messages, presenting challenges due to the absence of centralized administration \cite{quyoom2020security}. The Grayhole Attack, a variant of the Blackhole Attack, involves untrustworthy vehicles selectively forwarding certain data packets while dropping others without detection \cite{sheikh2019survey}.
  
Preserving confidentiality in VANETs involves using certificates and shared public keys for message encryption, ensuring access only to designated vehicles. However, several common threats to confidentiality exist. Eavesdropping, a prevalent attack on wireless networks like VANETs, focuses on obtaining unauthorized access to confidential data. This passive attack compromises confidentiality without having an immediate impact on the network, allowing the collection of useful information, including location data, which could be exploited for tracking vehicles. Another threat is traffic analysis in VANETs, which poses a serious risk to confidentiality and privacy as attackers track transmitted messages, analyze their content, and extract valuable information \cite{jan2021survey}. In a man-in-the-middle attack during V2V communication, the attacker inspects and alters messages, gaining control over the communication while the entities believe they are communicating directly in private. Additionally, a social attack aims to distract drivers by sending immoral and unethical messages, impacting the driving experience and the performance of the vehicle within the VANET system \cite{hamdi2021review}.
 
Authentication plays a crucial role in defending VANETs against attacks from both internal and external malicious nodes. Various threats and attacks target VANET authentication. Sybil attacks, introduced in \cite{douceur2002sybil}, involve sending fake messages with multiple identities, disrupting normal operations, and causing perceived traffic congestion in VANETs \cite{safwat2022survey}. GPS spoofing disrupts vehicle navigation by generating false information and misleading other vehicles with deceptive GPS locations. A tunneling attack in VANETs, similar to a wormhole attack, involves creating an additional communication channel, known as a tunnel, between distant parts of the network, enabling far-reaching nodes to communicate as neighbors. Replay attacks, where valid data is deceitfully retransmitted to create unauthorized and malicious effects, pose a prevalent security threat in VANETs. Additionally, the key and/or certificate replication attack encompasses the deceptive use of duplicated keys and/or certificates from other vehicles, complicating authentication and identification processes for trusted traffic authorities, especially during disputes \cite{hamdi2021integrity}.\\

Data integrity, ensuring that transmitted data remains unchanged, is a vital aspect in VANETs, and several threats target data integrity. In a masquerading attack, intruders use registered user IDs and passwords to infiltrate the system, disseminating false or malicious messages that appear to originate from the registered node. Exploiting each node's unique MAC address and IP address, masquerading attackers utilize stolen valid credentials for deceptive broadcasts within the VANET network. The replay attack repeats or delays transmission by injecting previously valid data into the VANET network, posing challenges for traffic authorities in identifying vehicles during emergencies \cite{safwat2022survey}. The illusion attack involves generating false data to deceive the VANET network, undermining integrity and data trust. Attackers exploit existing road conditions to create illusions for nearby vehicles, diminishing system performance through deceptive traffic warnings and undesirable bandwidth consumption. In a message tampering attack, assailants manipulate recent message data before transmission, altering information during congestion to mislead users into changing their driving paths. Attackers may delete, alter, or create new messages to achieve the intended purpose of the attack \cite{azees2016comprehensive}.

Ensuring nonrepudiation in communication means that both the sender and receiver cannot deny the transmission and receipt of messages during a dispute. Repudiation, a malicious action, occurs when an attacker disrupts the transmission or reception of critical messages, causing severe damage. In this attack, the assailant denies involvement in sending and receiving messages, particularly during disputes \cite{safwat2022survey,sheikh2019survey}.



\begin{figure}[H]
  \centering
    \includegraphics[width=0.8\textwidth]{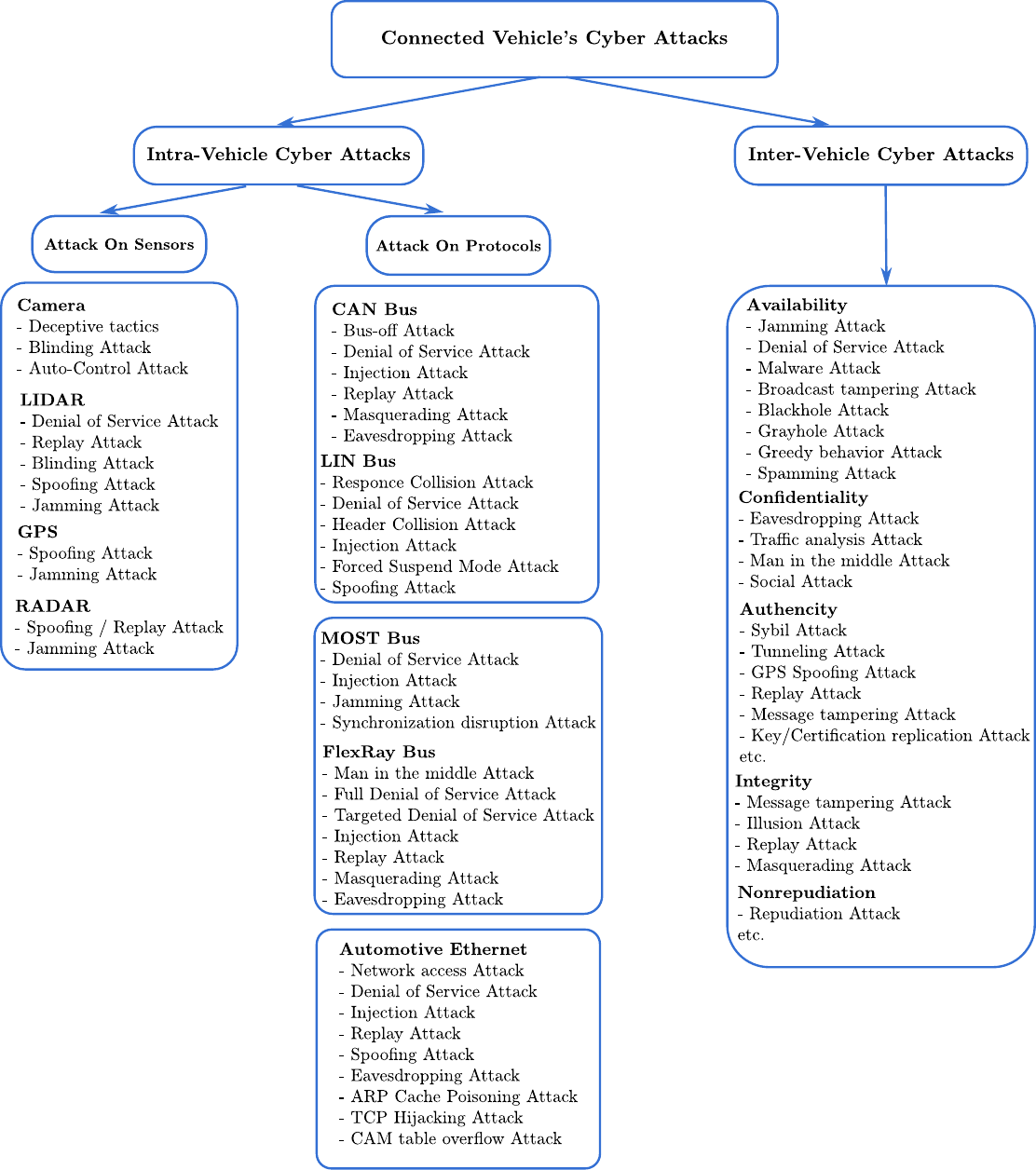}
    \caption{Toxonomy of connected vehicles under cyber attack}
    \label{fig:CVs attack}
    \end{figure}

To provide a visual overview, Figure \ref{fig:CVs attack} presents a taxonomy of connected vehicle cyber attacks. This analysis aims to shed light on the intricate landscape of cybersecurity challenges CVs face, offering valuable insights for safeguarding the future of connected automobiles.

Zero Trust provides a proactive model to tackling these challenges by implementing its principles within vehicular networks. Through the adoption of a ZT architecture, vehicles can authenticate all communication participants, whether they originate from within the vehicle itself or external entities, thereby mitigating the risks posed by cyber-attacks. Moreover, ZT addresses trust management challenges by continuously verifying and authorizing access to resources, regardless of the location of the vehicle or network connection status. This integration of ZT principles establishes a comprehensive security model that safeguards both intra- and inter-vehicular communication channels, effectively managing trust dynamics in dynamic vehicular environments.

\label{sect: CVs Attack}



\section{Zero Trust Models for Enhanced Connected Vehicle Security}

The emergence of connected vehicle technologies and zero trust cybersecurity paradigms catalyzes a paradigm shift in automotive safety protocols. The zero trust model, predicated on the axiom of continuous verification for all system interactions, comes as an indispensable bulwark against the growing number of cyber threats in the increasingly interconnected vehicular system. The integration of zero trust architecture into connected vehicle systems provides robust defence mechanisms, significantly enhancing the overall cybersecurity posture and consequently fostering safer and more secure automotive experiences. 

In this section, we will initially examine some of the existing security solutions relevant to mitigating cyber-attacks on connected vehicles, as well as the trust models related to them. Furthermore, we delve into the primary focus of this paper, which is a comprehensive analysis of zero trust security in CVs. We analyze the existing literature in this field, presenting each approach and its advantages.

Blockchain technology, with its distributed ledger and cryptographic features, presents a groundbreaking solution for revolutionizing data management in self-driving cars and CVs. By enabling real-time traffic analysis, accident reduction, optimal route identification, and minimized travel time, blockchain overcomes challenges like RADAR interference and regulatory gaps. Its tamper-proof, decentralized storage enhances data security and integrity, ensuring decentralized security, transparency, and traceability in CVs communications \cite{ahmad2024machine}. Concurrently, machine learning approaches, encompassing both supervised and unsupervised learning, play a crucial role in safeguarding VANETs from cyber threats. Techniques such as Naive Bayes, decision trees, and random forests in supervised learning effectively counter various attack types, including location falsification attacks. Unsupervised learning, particularly the K-means algorithm, detects jamming attacks by distinguishing benign interference from intentional jamming, providing valuable insights into data without prior knowledge \cite{hankins2023eyes}. 

Moreover, a zero trust architecture for connected vehicles emerges as a secure and scalable solution, prioritizing the security and privacy of car owners and passengers. By addressing challenges associated with traditional centralized systems and reducing risks associated with the Internet of Things, this model recommends the vast adoption of CVs technologies. Combining the ZT model with new solutions such as machine learning, deep learning, and blockchain improves cyber attack risk mitigation for CVs, as documented in future investigations.

Table 4 summarizes various ZT techniques in the context of vehicular systems. It encompasses fields such as electric vehicle infrastructure, intelligent connected vehicles (ICVs), the Internet of Connected Vehicles (IoCV), connected and autonomous vehicles, and Zero Trust techniques in vehicular networks.

\subsection{Trust models in Connected Vehicle Systems}
Trust models are essential for managing secure and dependable networks among vehicular nodes. These models assess the reliability and legitimacy of messages circulated within VANET system. Three key trust models include Entity-based Trust Models (ETM), which foster trust relationships in VANET entities; Data-based Trust Models (DTM), which assess data reliability among entities; and the Hybrid Trust Model (HTM), which integrates both ETM and DTM for a comprehensive approach to trust evaluation.

\subsubsection{Entity-based Trust Models}
    
The Entity-based Trust Model (ETM) assesses vehicle credibility through direct and recommendation-based trust, considering interactions among vehicles. This system calculates trust values for individual vehicles, identifying untrusted or malicious ones. It employs a distributed approach, with vehicles autonomously computing trust values. Enhancements like vehicle clustering and RSU-based trust computation are integrated to increase accuracy and reduce computational overhead.

There are limited traditional trust models proposed in the literature. \textit{ Marmol and Perez.} \cite{marmol2012trip} proposed a trust and reputation-based trust model (TRIP) to identify malicious or selfish nodes that transmit deceptive messages over the network rapidly and effectively. 
In \cite{xia2019attack}, a novel trust inference model was proposed. This model measures the trust level of a certain vehicle by combining subjective trust and recommended trust. The ETM exhibits strong resistance to a variety of attacks, including bad-mouthing attacks and collusion attacks. Nonetheless, scientists have proposed future work that includes considerations for deployment areas, network applications, and security levels, all of which may reveal potential coverage gaps. Trust computations as targets highlight a significant security challenge, and emphasis on defense mechanisms underscores potential vulnerabilities. 

A study by Minhas et al. \cite{minhas2010multifaceted} suggested a multifaceted trust model for VANETs that includes role-based, experience-based, priority-based, and majority-based trust to help build trust in environments that are always changing. The model is effective in countering deceptive information from malicious agents. In another ETM proposed in the literature by \cite{cui2019rsma}, the researchers introduced a Reputation System-based Lightweight Message Authentication (RMSA) model for 5G-enabled vehicular networks, addressing deficiencies in traditional PKI-based authentication. The model managed by a Trusted Authority (TA), restricts vehicles with reputation scores below a threshold from participating in communication, reducing untrusted messages. Built on the Elliptic Curve Cryptosystem (ECC) and supporting batch authentication, the scheme demonstrated security against various attacks. 
    
In \cite{hu2016replace}, scholars recommended a trust-based platoon service recommendation scheme, REPLACE, designed to address challenges in ITS, particularly within vehicular platooning systems. In REPLACE, a reputation system collects and models user vehicle feedback to help identify poorly behaved platoon head vehicles. An iterative filtering algorithm is employed to handle untruthful feedback. The proposed scheme is shown to be secure and robust against various attacks. Researchers in \cite{li2013rgte} outlined the Reputation-based Global Trust Establishment scheme (RGTEs) as a solution to the challenges faced by existing trust establishment methods in VANETs, particularly in rapidly changing environments. RGTEs utilize statistical laws to safely share trust information in VANETs, , thereby improving trust establishment efficiency and accuracy. To detect and identify bad nodes, the system dynamically adjusts thresholds based on real-time reputation status. 
    
Overall, the entity-based model faces difficulty collecting adequate trust evidence when vehicles join VANETs with limited interaction information \cite{wang2022decentralized}. Similarly, the ETM grapples with data sparsity, particularly in scenarios where vehicles are widely dispersed, impacting the accuracy of trust evaluation \cite{zhang2023introduction}.

\subsubsection{Data-based Trust Models}

In order to evaluate the reliability of communications, the data-based trust model (DTM) gathers and filters information from various sources, including neighbors and RSUs. This is particularly crucial in IoVs since messages usually contain much information. Because the messages are time-sensitive, the data-based trust model computes real-world information per message, such as traffic conditions and position data. 

\textit{Zaidi et al.}\cite{zaidi2014data} proposed a cooperative detection and Correction (C-DAC) scheme to strengthen VANET security by stopping rogue nodes from sending false emergency messages. This data-driven approach does not require infrastructure or revocation lists because it utilizes cooperative data flow among vehicles to identify and correct rogue nodes. The method uses a network model to allow each node to independently predict its current state to guarantee fault tolerance and resistance against false data injection in VANETs. The scheme faces some limitations in identifying actual rogue nodes during Sybil attacks, where a rogue node presents multiple identities. Additionally, coordinated attacks by multiple attackers blocking lanes and sending false messages with different identities may go undetected by C-DAC. The technique struggles to distinguish between rogue and honest nodes during sudden density increases in the case of an actual accident, making it challenging to identify rogue nodes during this transition phase.

Another DTM proposed in \cite{shaikh2013trust} is a decentralized and privacy-aware trust management scheme. Addressing challenges posed by the high mobility and the temporary nature of the VANETs, their method ensures user privacy, robustness against security threats, and linear time complexity for real-time implementation. RaBTM, as an RSU-aided beacon-based trust management system for VANETs, was introduced by authors in \cite{wei2012efficient}. The primary objectives of RaBTM are to enhance the accuracy and reduce the delay in making trustworthiness decisions, thereby improving safety and location privacy for vehicles. Similar to scientists in \cite{wu2011rate}, they proposed an RSU-aided scheme for data-centric trust establishment in VANET called RATE; this model can efficiently conduct trust establishment. Thus, RATE leads to increased delays compared to distributed mechanisms. Some factors influenced the performance of RATE, such as observing conditions, event evolving status, and the proportion of malicious data.\\
\textit{Raya et al.}\cite{raya2008data} presents a data-centric trust framework for mobile ad hoc networks, addressing limitations in traditional entity-based models. It computes trust for individual data, combines related data, and employs Bayesian inference and dempster-shafer theory for validity inference. Evaluation in vehicular networks demonstrates resilience to attacks, stable convergence, and effectiveness in specific scenarios, meeting life-critical network requirements.\\
The drawbacks of DTMs encompass issues such as latency and data sparsity. Latency arises due to the abundance of data from various sources, leading to the potential dismissal of valuable information or overwhelming significant data \cite{khan2021security}. Another limitation is the inability to establish a long-term trust relationship among vehicles. Instead, data-based models can only establish short-term trust for received data based on individual events. This restriction necessitates repetitively constructing trust relationships for each event without utilizing previous trust data. Additionally, accurately judging the reliability of messages becomes challenging when the quantity of received messages is insufficient\cite{wang2022decentralized}.

\subsubsection{Hybrid Trust Models}
Hybrid trust models (HTMs) evaluate vehicle confidence and data reliability in this category, integrating entity-based and data-based assessments. The HTMs calculate trust values for both vehicles and messages, with the latter often based on entity trust. However, the hybrid approach can be more complex than other models and introduce higher communication overheads. This method efficiently considers both entity and data trustworthiness in evaluations, ensuring that data assessed by trusted entities is perceived as trustworthy by other network nodes in VANETs.
 
The research in \cite{wei2014adaptive} addressed decision-making challenges in VANETs, focusing on the accuracy and delay of received event messages. They presented a model for adaptive decision-making using RSUs for rapid and efficient support. Compared to a baseline, simulations using a range of trust attacks showed that the suggested approach increased detection accuracy and decreased decision delay. The paper tackles the crucial problem of erroneous traffic warning signals in VANETs, stressing their possible influence on the decisions of the driver, time, and fuel efficiency. An Event-based Reputation System (ERS) was developed in \cite{lo2009reputation}, for drivers, the system includes a dynamic reputation evaluation mechanism that determines the importance and reliability of incoming traffic messages. Results show that the system effectively stops the spread of false messages in different VANET contexts when a proper reputation adaption mechanism and threshold settings are used.  
 
The research covered in  \cite{chen2013beacon} focused on managing trust in VANETs, with a specific focus on protecting location privacy. The authors suggested a Beacon-based Trust Management system (BTM) to thwart internal attackers trying to deliver misleading communications in VANETs with greater privacy. To assess the dependability and efficiency of the system. Simulations involving message suppression, fraudulent messages, and altering assaults were carried out. The researchers come to the conclusion that the BTM system is workable for location privacy-enhanced schemes in addition to being robust against trust attack models. 

\textit{Placzek et al.}\cite{placzek2016detection} examined how Sybil attacks affect VANET performance in traffic light management applications. They proposed a technique to identify malicious data injected by vehicle nodes who participate in Sybil attacks with position verification combined with a driving behavior model. Simulation studies of a decentralized, self-organizing system controlling traffic signals at various intersections in an urban road network demonstrate the effectiveness of the approach.

Another HTM suggested by scholars in \cite{li2015art} is an Attack-Resistant Trust model (ART) for VANETs to enhance the trustworthiness of both traffic data and vehicle nodes. The ART scheme assesses data trust, determines the trustworthiness of reported traffic data, and assesses node trust, evaluating the reliability of nodes in VANETs. The two functional dimensions of trust evaluation are functional trust and recommendation trust, which reflect the capacity of any node to carry out its functionality and the reliability of suggestions, respectively, made towards other nodes. The ART scheme resists a wide range of malicious attacks; hence, the scheme is appropriate for various applications in VANET, including mobility, environment protection, and traffic safety.

Increased system overhead characterizes the hybrid trust model because it inherits complications from entity-based and data-based approaches. The complexity of the model, based on the establishment of adequate trust relationships between vehicles and the evaluation of the dependability of every message data, increases resource consumption and computational demands. Besides, limited data poses challenges to the model \cite{wang2022decentralized}, which could challenge trust assessment accuracy. Despite the benefits of considering entities and data, the HTM's increased complexity and system overhead present significant drawbacks. However, these drawbacks typically arise when dealing with small data sets.
\\

\subsection{Zero Trust in Electric Vehicles Infrastructure} 
Electric vehicles (EVs) have many advantages, including reducing emissions and operation expenses. Nevertheless, like any technology, they also have weaknesses that should be considered. Such risks are going on in the development of EV charging infrastructure. Public charging stations share risks like tampering or hacking that can severely impair the availability and reliability of the charging service. Due to the rapid digitization of EV chargers, \textit{Li et al.} \cite{li2023zero} have shown that these devices are becoming increasingly vulnerable. Hence, it proposes a blockchain and zero trust architecture-based security scheme to protect those chargers and the cloud platforms. This work is proposed to save EV security, energy crises, and environmental issues despite being proven effective during attacks. Another research on EV by \cite{zhao2023research} suggests a security model for smart charging stations, using ZT model, Elliptic Curve Cryptography, and access control to secure vehicle-node communication. This method targets security challenges, adapting to the construction needs of the station and paving the way for secure operation.  

 \subsection{Zero Trust in Intelligent Connected Vehicles}
Intelligent Connected Vehicles (ICVs) integrate modern technologies like Artificial Intelligence, Big Data, and Cloud Computing alongside simple connections such as V2V and V2I for smart driving functions related to navigation and predictive maintenance. Authors in \cite{huang2023overview} investigate the application of the zero trust model to address challenges in ICV platoons, focusing on dynamic behaviors, information interactions, and security threats. It explores the ZT model, focusing on the processes of authorization and confirmation among connected nodes within a group. The ZT model discussion encompasses information perception, full information chain communication, and fault-tolerant control techniques in ICV systems. The paper defined the ZT theory, used mathematical models to describe the ICV system, and discussed impact factors such as system dynamics and communication topology. \textit{Cui et al.} \cite{cui2023trust} discuss trust assessment of communication topology nodes in ICVs running in a zero trust environment. A customized trust model is proposed that calculates the trust function of nodes with time based on weighted calculations. The work establishes the communication topology model in a ZT environment, evaluates the consistency of ICV queues, and analyses the impact of different node trust values on the whole topology.

The researchers in \cite{li2023adrc} presented a method using Active Disturbance Rejection Control (ADRC) to reject the influence of unknown input disturbances, which occur due to the ZT environment, on the ICV queuing system. The focus was not only on disturbance rejection but also on ensuring the stability of vehicle formations and increasing the overall system robustness and security. In combining ZTA concepts and ADRC, the proposed model tried to strengthen the management of vehicle queues in ICVs, thus targeting issues emerging from dynamic driving scenarios and cyber threats. These authors carried out extensive simulations and analyses to test the efficiency of the proposed ADRC, which was proven to be effective in keeping the vehicle queue safe and efficient. On the other hand, scientists in \cite{zhang2023distributed} make an effort to address the problem of fault-tolerant control for the dynamic communication topology of ICV platoons in a ZT environment. Here, the communication structure is reconfigured by adding a trust factor, which opens up the path toward a fault-tolerant control policy. Here, a state observer and error estimator are used, which includes the cases where the system cannot get its state. The stability of the error system is studied using the Lyapunov method, while the distributed controller has been designed to wipe out errors.\\

\subsection{Zero Trust in Internet of Vehicles}

 \textit{Fang et al.} \cite{fang2022zero} proposed an IoV security scheme that integrates MFA, blockchain encryption, and a ZTA to fortify data transmission, user privacy, and system integrity. Although it faces limitations due to a lack of deployed vehicle nodes for comprehensive simulation, IoV was also of interest in \cite{song2022new}. The research introduces a two-step authentication system for modern car rental services. It starts with a secure smart key generation process using static authentication and then implements continuous driver authentication while driving, utilizing fingerprint, NFC, and facial recognition data. Subsequent research in \cite{wang2023attribute} proposes an Attribute and User Trust Score-based Zero Trust Access Control Model (AU-ZTAC) to address the intricate challenges in IoV security. This model combines the zero trust and attribute-based access control models, providing fine-grained dynamic access control. The methodology integrates trust evaluation into the access control process, ensuring a comprehensive reflection of the intent of the user.
 
Researchers in \cite{fang2024decentralized} developed a decentralized handover authentication scheme for seamless authentication in zero trust IoV networks. It involves multiple Authentication Cooperators (ACs) at the network edge collecting device/location-related features of vehicles for identity verification. During vehicle movement, Access Points (APs) select new ACs for handover authentication, ensuring minimal disruption. They also introduced a situation-aware AC selection and update algorithm and proposed a hierarchical blockchain-assisted mechanism for secure information transfer and reputation management. Their scheme outperforms existing methods in terms of authentication accuracy and handover time, offering continuous IoV protection through collaborative authentication and trajectory prediction.

Furthermore, an IoV-ZT conceptual model architecture was designed by \textit{Zhang et al.} \cite{zhang2023based} to address the security problems in IoV. It offers the Linkable Ring Signature Map Review (LRSMR) scheme for privacy security under autonomous vehicle map review scenarios. Based on a linkable ring signature mechanism, it utilizes the SM2 digital signature algorithm. LRSMR supports the preservation of privacy for user identity and signature linkability within a ring, enhances the reliability of review data through user credit demand, and provides rewards for the distribution of incentives, which will provide an incentive for the participation of users. The security analysis proved that LRSMR is correct, unforgeable, unconditionally anonymous, linkable, and nonslanderous. According to simulation results, LRSMR has demonstrated high efficiency in both communication overhead and computation costs.

\textit{Zayed et al.} \cite{zayed2022owner} addresses security challenges within the Internet of Connected Vehicles (IoCV) by proposing the implementation of ZTA to enhance information exchange and security measures. To achieve this, the methodology focuses on verifying the identity of vehicle owners, utilizing techniques such as recognizing license plates and retrieving owner details. 

Another work in the context of IoCV proposed in \cite{wang2023distributed} involves researchers exploring distributed fault detection in a ZT environment. It studies the influence of the zero trust paradigm on fault diagnosis and analyses the stability conditions of the £2 observer.

\subsection{Zero Trust in Connected and Autonomous Vehicles}

\textit{Anderson et al.} \cite{anderson2023zero} add to this research landscape an aim to enhance the security of connected and autonomous vehicles. Their goal is to identify vulnerabilities in controller area network bus technology. The authors propose the ZTA to protect CAV sensors and control networks against cyber threats. The investigation examines the existing attacks against CAN networks and solutions to enhance the security posture. The authors propose ZTA in CAVs that address the problems of fabrication, suspension, and masquerade by requiring all the ECU nodes to register with the ZTA controller for contextual checks and permission for the legitimate flow at the CAN bus. 

\textit{John Blåberg Kristoffersson} \cite{blaaberg2022zero}, in his research, presents the investigation of the implementation of a zero trust network in the automotive ethernet for fully autonomous vehicles. They emphasize the shift in security from the perimeter to the ZT model. It examines security protocol performance, suggests "Bump-in-the-Wire" devices for nodes lacking security, and explores key distribution using 802.1X authentication with RADIUS and certificates within automotive Ethernet. 

The researchers in \cite{kondaveety2022zero} investigated the security risks associated with connected and autonomous vehicles, which rely on sensor data for decision-making. They proposed a security architecture based on the ZT model, focusing on maintaining confidentiality, integrity, and data availability.

To assess the effectiveness of their proposal, they conducted simulations using the NeSSi2 network simulation tool and analyzed captured packets with wire sharks. They specifically simulated common cyberattacks like DDOS and packet sniffing to compare the security performance of the proposed architecture with traditional approaches. Results showed that the ZTA offered superior protection against these attacks, highlighting its potential for enhancing security in next-generation automobiles.

\textit{Sullivan et al.} \cite{sullivan2024observe}  developed OBSERVE, a ZT protocol for CAVs to endorse each other's trajectories in real-time to prevent the potential malicious or
false data for navigation and path planning decisions, using blockchain. This model allows vehicles to self-organize and reach a consensus on the truthfulness of nearby trajectories of vehicles through a consensus mechanism. They employed simple machine learning algorithms for trajectory prediction to reduce computational overhead and energy consumption. OBSERVE was validated on realistic data, using simulated connected vehicle trajectories from New York City, demonstrating higher prediction accuracy with lower computation overhead.

\subsection{Other Zero Trust Models}

 The researchers in \cite{fowler2023practical} implemented a Free Space Optical Quantum Key Distribution (FSO-QKD) system within a V2I application as part of the AirQKD project. To enable secure communication, they developed various technologies, including single-photon components, FSO systems, QRNGs, and cryptographic protocols. Through a practical experiment, the researchers demonstrated the functionality of the QKD system in generating symmetric keys for a zero trust security scheme within V2I applications. Their goal was to replace certificate-based PKI systems with QKD-generated symmetric keys, contributing to the advancement of 6G communications.
\textit{Liu et al.} \cite{liu2022blockchain} developed a blockchain-based information-sharing protocol for ZT environments, focusing on IoT nodes like autonomous vehicle sensors. This protocol facilitated autonomous data exchange without centralized servers and included authentication and consensus mechanisms. Roadside units, smart contracts, vehicles, shared files, and blockchain were all part of the system that ensured blockchain-based identity verification was used before sharing current road information. It packed blocks with the best reputation, using RSUs and smart contracts to drive the consensus method value after a predetermined quantity of exchanges. Through vehicle IDs and signatures, it ensured non-repudiation, and a bonus system encouraged proper information sharing. Additionally, based on reputation values, it gave users better rating priority throughout subsequent sharing processes.

A trustworthy Ultra-Reliable Low Latency Communication (URLLC) resource slicing and scheduling method for 6G zero trust vehicle networks has been suggested by the authors in \cite{hao2021urllc}. Using edge, convergence, and cloud servers as the three infrastructure layers, their method incorporates a reputation mechanism. They developed a subjective logic model for reputation monitoring of access points in the network, enhancing the security degree of accessed nodes and trustworthiness in the ZTA. Furthermore, they created a federated asynchronous reinforcement learning algorithm for optimizing the deployment of slice resources while protecting the privacy of vehicles and optimizing efficiency.

\newpage


\begin{longtable}[c]{|m{2.75cm}|m{0.8cm}|m{3.5cm}|m{3.5cm}|m{4cm}|} \caption{Zero Trust model for Connected  Vehicles} \\
\hline
\textbf{Reference}  & \centering \textbf{Year} & \centering \textbf{Focus} & \centering \textbf{Methodology} & \textbf{Benefits} \\ 
\hline
\textit{ Li et al.}\cite{li2023zero}    &  2023 & Securing EV chargers and cloud platforms  &  Blockchain and zero trust principles Security scheme  & Protection of EV security, addressing energy crises and environmental issues \\
\hline
\textit{Zhao et al.} \cite{zhao2023research} & 2023 & Securing  vehicle-node communication at charging stations &  Zero trust model, Elliptic Curve Cryptography, access control & Secure operation of smart charging stations, adaptation to construction needs \\
\hline
\textit{Zayed et al.} \cite{zayed2022owner}& 2022 & Implementing Zero Trust Architecture in IoCVs & Verifying the identity of vehicle owners, recognizing license plates, retrieving owner details & Enhanced information exchange, improved security measures  \\
\hline
\textit{Wang et al.} \cite{wang2023distributed} & 2023 & Exploring fault detection in IoCVs under zero trust environment &  Investigation of zero trust paradigm impact, analysis of £2 observer stability conditions & Exploration of fault diagnosis, analysis of stability conditions\\
\hline
\textit{Zhang et al.} \cite{zhang2023distributed} & 2023 &  Tackling fault-tolerant control in dynamic communication topology of IoCV platoons &  Introduction of trust factor, design of fault-tolerant control model & Reshaping communication structure, designing fault-tolerant control model \\
    \hline 
\textit{Bao et al.} \cite{bao2023stability} & 2023 & Investigating internal stability of vehicle platoons in zero trust environment & Establishment of linearized vehicle dynamics model, design of distributed control law & Analysis of internal stability conditions of vehicle platoons  \\
\hline 
\textit{ Fang et al.} \cite{fang2022zero} & 2022 &  Data transmission, user privacy, and system integrity in IoV environment & Integration of multifactor authentication, blockchain encryption, and zero trust architecture & Improved data transmission, enhanced user privacy and system integrity\\
\hline 
\textit{ Song et al.} \cite{song2022new} & 2022 & Introduction of two-step authentication system for modern car rental services & Implementation of the secure smart key generation process, continuous driver authentication & Improved security for car rental services \\
\hline 
\textit{ Wang et al.} \cite{wang2023attribute} & 2023 & Addressing IoV security challenges & Integration of zero trust and attribute-based access control model, trust evaluation integration & Dynamic access control, comprehensive reflection of the intent of the user\\ 
\hline 
\textit{ Fowler et al .}\cite{fowler2023practical} & 2023 & Implementation of FSO-QKD system to secure communication in V2I applications of the  AirQKD project &  Developement of various technologies, including single-photon components, FSO systems, QRNGs, and cryptographic protocols &  Functionality demonstration of QKD system, generation of symmetric keys \\
\hline 
\textit{ Cui et al.}\cite{cui2023trust} & 2023 & Evaluating trust in communication topology nodes in ICVs operating in zero trust environment &  Development of a tailored trust model using weighted calculations  &  Evaluation of node trust levels, analysis of topology consistency \\
\hline
\textit{ Anderson et al.} \cite{anderson2023zero} & 2023 & Identifying vulnerabilities within CAN bus technology in CAVs  & Examination of attack methods, proposed solutions for security enhancement & Provides robust security by detecting and preventing various attacks, ensuring continuous context checks and flow authorization on the CAN bus\\
\hline 
 \textit{Kristoffersson.} \cite{blaaberg2022zero} & 2022 & Evaluation of zero trust network implementation in automotive ethernet for autonomous vehicles & Performance evaluation, suggestion "Bump-in-the-Wire" of security measures & Provides efficient and low-resource security enhancement with MACsec encryption, particularly demonstrated by its performance surpassing IPsec in latency without significant resource expenditure.\\
  \hline 
 \textit{ Huang et al.} \cite{huang2023overview} & 2023 &  Investigating the application of zero trust model in ICV platoons &  Analysis of information perception, communication topology, and fault-tolerant control &  Exploration of information perception, full information chain communication, fault-tolerant control mechanisms \\
 \hline 
\textit{Fang et al.} \cite{fang2024decentralized} & 2024 &  Authentication in zero trust IoV networks &  Decentralized handover authentication scheme for zero trust IoV networks &  Enhanced authentication accuracy, reduced time costs, and continuous protection for Internet of Vehicles (IoV) systems \\ 
\hline
 \textit{Kondaveety et al.} \cite{kondaveety2022zero} & 2022 & Security risks associated with connected and autonomous vehicles & Security architecture based on the Zero trust model & Provides superior defense against attacks, thereby enhancing security in future automotive systems\\
\hline
\textit{Sullivan et al.}\cite{sullivan2024observe} & 2024 & Verification the truthfulness of trajectories and other data shared by neighboring vehicles & OBSERVE a Blockchain-based Zero Trust Security
Protocol for CAVs & Higher prediction accuracy with lower computation overhead\\
\hline
\textit{Zhang et al.} \cite{zhang2023based} & 2023 & Security challenges in IoV & LRSMR an IoV-ZT conceptual model & Offers improved user privacy, enhanced data reliability, and increased participation, contributing to a more trustworthy and efficient map review process for autonomous vehicles\\
\hline 
\textit{Li et al.}\cite{li2023adrc} & 2023 & Impact of unknown input perturbations arising from the zero trust environment on the ICV queuing system & Active disturbance rejection control (ADRC) integrating ZTA principles & Ability to maintain a safe and efficient vehicle queue in dynamic driving environments\\
\hline
\textit{Liu et al.} \cite{liu2022blockchain} & 2022 &  IoT nodes like autonomous vehicle sensors & Blockchain-based information-sharing protocol for zero trust environments & The model prioritized users with higher ratings for future sharing processes based on their reputation values\\
\hline 
\textit{Hao et al.} \cite{hao2021urllc} & 2021 &  Security and efficiency challenges posed by zero trust vehicular networks in the context of 6G & Trustworthy Ultra-Reliable Low Latency Communication (URLLC) resource slicing and scheduling approach  & Efficiently scheduling slice resources and protecting vehicle information security\\
\hline
\end{longtable}


\section{Open problems and challenges}

\begin{figure}[H]
  \centering
    \includegraphics[width=0.8\textwidth]{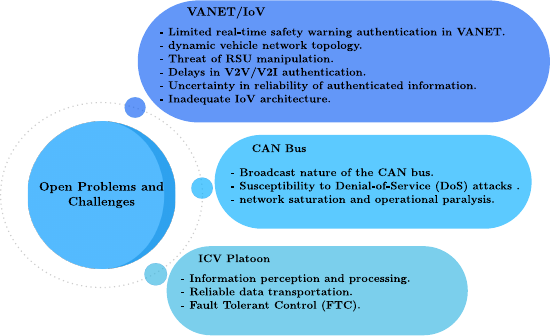}
    \caption{Open problem and challenges of connected vehicles under the zero trust model}
    \label{fig:problem15}
    \end{figure}

In this section, we will outline the potential challenges within the CVs communication system under the zero trust model, building upon the approaches presented in section 5, as summarized in Figure \ref{fig:problem15}. However, before delving into these specific challenges, it is essential to introduce the broader zero trust challenges. This will provide a comprehensive understanding of the overarching issues facing the implementation of zero trust principles in various technological environments.

Vendor lock-in poses a persistent challenge in the technology industry, where organizations face risks tied to over-reliance on specific vendors. This issue spans diverse sectors, including cloud computing, IoT, and emerging concepts like ZTA. In the early stages, ZT solutions raise concerns about potential lock-in as organizations assemble components from various vendors. Navigating this landscape requires standards for interoperability and strategic planning to avoid exclusive vendor dependencies, emphasizing the need for flexible and adaptable technological environments \cite{teerakanok2021migrating,phiayura2023comprehensive,khan2023zero}. 

Implementing ZTA in enterprises faces significant challenges due to the absence of industry standards providing clear guidance. This deficiency makes it difficult to ascertain the successful implementation of ZTA within organizations. Moreover, certain components of ZT platforms, like the PDP, lack uniform standards for information exchange. While some standards exist for sharing threat intelligence information, there is an overall lack of common standards in crucial areas. For instance, the decision-making process of ZTA carried out by the PDP relies on information from various sources, contributing to the complexity. In scenarios where a quick replacement is needed, encountering security or technical issues becomes a considerable challenge without established common standards, potentially leading to high costs \cite{teerakanok2021migrating,phiayura2023comprehensive}.

Minimizing disruptions to users is a significant challenge when deploying ZTA. When initiating the migration cycle, an enterprise introduces a new ZTA-based model in the parameterized zone to replace the legacy system, encouraging users to adopt the new system. Simultaneously, technical restrictions are slowly added to the old method to encourage users to switch to the ZTA-based system. As more users use the new system without issues, the enterprise considers moving the new workflow to the unprivileged network. However, technical issues and user disturbances may arise at any point during the transfer process, therefore remediation plans must be properly thought out. \cite{phiayura2023comprehensive}. 

Managing unmanaged devices and allocating resources are the two main implementation challenges of ZTA. ZTA would require enormous financial, technological, and human resource pools for its successful implementation. As a result, allocating resources becomes crucial to the model of ZTA implementation. In addition, there is always a problem with unmanaged devices. Maintaining control and oversight over the devices may present challenges for the organization, particularly in BYOD contexts. Privacy concerns make matters worse since the company is not allowed to monitor, control, or install the necessary software on the devices of the employees.
ZTA adoption requires a difficult balancing act between smart resource allocation and resolving the complexities of unmanaged devices in order to be successful.
 \cite{khan2023zero,phiayura2023comprehensive,teerakanok2021migrating}. 

Making the right choices about access requests mostly depends on the TA or the cognitive process that powers the PE of the ZTA. Because of its dynamic character, TA is a dynamic decision mechanism that uses machine learning to gradually increase its decision-making capacity. This dynamic is crucial because a false positive or false negative result from TA could raise security concerns by granting unauthorized people access or by permitting malicious access. Continuous observation and fine-tuning of TA parameters are essential to ensure accurate functionality. Moreover, the integration and collaboration of supported systems within the ZT platform are emphasized, preventing isolation to maintain efficient visibility and analytics. The normalization, filtering, and correlation of information from diverse sources further enhance the overall efficiency of the TA in the ZTA environment \cite{phiayura2023comprehensive,teerakanok2021migrating}.

\subsection{Zero trust challenges in Vehicular Networks}
Implementing zero trust algorithms in VANET faces challenges due to the incapacity of existing solutions to authenticate real-time safety warnings from vehicles and the limited deployment capabilities of OBUs. The dynamic network topology resulting from vehicle movement further complicates adaptation. Despite these limitations, there is potential for leveraging ZT principles in securing background communications with backend servers and RSUs, offering a more focused application of these concepts in vehicular environments\cite{sateesh2020state}. Otherwise, in ZT vehicular networks, there is a potential threat of attackers maliciously manipulating RSUs, emphasizing the need to safeguard against such risks. Leveraging machine learning and deep learning algorithms can enhance authentication and security. These technologies enable real-time communication analysis, improving the accuracy of safety warnings and detecting malicious activity. 

The need for authentication and authorization before each V2I or V2V communication can lead to significant computational energy consumption and transmission delays. This may impact the fulfillment of service requirements and business demands. Additionally, even after passing identity verification, the reliability of uploaded information remains uncertain, posing a risk to transportation safety with the potential for invalid or false information\cite{hao2024exploiting}. Blockchain can enhance security by providing decentralized authentication mechanisms and enforcing access control policies through smart contracts. 

\textit{Fang et al.}\cite{fang2022zero} their implementation scheme within the ZT model faces a notable limitation due to the inadequate representation of the IoV architecture. The insufficient deployment of vehicle nodes in the simulation hinders the ability of the scheme to mimic the complexities and scale of real-world IoV scenarios accurately.

In a ZT environment, the default model is to not inherently trust any entity and verify the trustworthiness of each interaction. As mentioned in \cite{wang2023distributed}, Real-time trust updates become crucial in dynamic scenarios where conditions, behaviors, or vehicle relationships are constantly changing.

The challenge encountered by researchers \cite{anderson2023zero} in the ZTA for CAVs concerning DoS attack vectors on the CAN bus. The broadcast nature of the CAN bus makes it susceptible to DoS attacks, allowing attackers to saturate the bus and render the entire network unusable. This vulnerability is attributed to the exclusive reliance of architecture on the CAN bus as the communication channel. The potential consequences of a DoS attack include the necessity for the CAV to cease operation to ensure occupant safety, as the ZTA cannot process control or response messages during such an attack, resulting in a paralyzed network. Implementing security measures in safety-critical real-time systems is challenging due to resource limitations. Intrusion Detection Systems (IDS) for CAN networks have emerged to address these challenges. IDS can be signature-based, which relies on known attack patterns, or anomaly-based, which detects deviations from normal behavior. While signature-based IDS requires regular updates and may miss unknown attacks, anomaly-based IDS can detect previously unknown threats \cite{bozdal2020evaluation}. Implementing IDS on the CAN bus helps detect and respond to DoS attacks proactively, ensuring uninterrupted operation and safety in CAV environments.


\subsection{Zero Trust challenges in Intelligent Connected Vehicles } 

\noindent Implementing Zero Trust in ICVs platoons has led to several challenges and issues. According to authors in \cite{huang2023overview} there are three main challenges :

\begin{enumerate}
\renewcommand{\labelenumi}{\alph{enumi}}
\item \textit{. Information perception and processing challenges}

One challenge involves dynamic deception within the visual area of ICV platoons, demanding the identification and elimination of deception targets based on multisource feature information. This entails ensuring the concordance of features between multisensor data and multi-feature tag data that are correctly matched.
The other challenge is that malicious nodes cause an unbalanced match of decisions and information asymmetry in the ICV system. Solving this problem involves research on cooperation competition mechanism-driven approach, including the feature extraction mechanism of multi-dimensional 
data and lightweight distributed mechanism for cooperation competition games. Also, although promising, the cooperative driving of ICV platoons in real system implementations faces operational challenges.

    \item \textit{. Reliable data transportation challenges}
    
The problems in the data security and access control of ICV systems in the model of a ZT scenario span from the information-sensing and control terminals, communication infrastructure, and cloud platforms. With so many terminal device nodes, the high number itself contributes toward a detrimental effect on the process of data fusion, it will lead to further threats in forms of malicious attacks and unauthorized access. The shift from partial to total ZT communication among interaction nodes introduces challenges. These challenges encompass designing secure data encryption, transmission, and processing methods, considering periodic migration characteristics. Addressing confidentiality, authenticity, and integrity concerns in information chain transmission while accommodating the demand for weak centralization is vital. Furthermore, challenges include establishing adaptive forgetful access control for intelligently networked vehicle platoon data, covering global index setup, local index framing, credit authorization allocation, and implementing a risk-adaptive forgetful access control model.

\item \textit{. Fault Tolerant Control (FTC) challenges}

The transportation scenarios demand fault tolerance since the disruption of the ICV platoon might affect the entire platoon. The FTC challenges in ICVs platoons under the ZT model The comprehensive modeling of all the nodes of the ICV platoon needs model identification, either by model or data approaches. Independent controllers are required for each node. Second, the trust values need to be evaluated for all the nodes. The result would be to start an authorization process, and the rejected vehicles would be treated as external disturbances. Accepted vehicles need a distributed FTC controlling scheme and consistency strategy for ICV systems based on the perception of the environment. The data security transmission among the nodes of the ICV system is also a challenge. The necessary malicious attacks, false location transmission, or data tampering could present serious consequences for the entire ICV platoon. These challenges highlight the intricate nature of achieving fault tolerance in the context of ICV platoons under a ZT model.

Scholars also mentioned some extended issues in evaluating node trust values and network vulnerability within the ZT model. It highlights the need for trust evaluation mechanisms and authorization processes to assess the reliability of communication nodes, data sources, and control commands in the ICV system. This involves evaluating entity trustworthiness and granting access based on trust levels, following the core principle of "never trust, always verify." The document also explores challenges in heterogeneous vehicle platoons under the ZT model, emphasizing effective data perception in unstructured environments and multi-degree-of-freedom feature identification for diverse vehicles to enhance safety.

\end{enumerate}

\section{Conclusion}
In reshaping cybersecurity strategy, zero trust represents a significant departure from the conventional reliance on a trusted internal network. This model acknowledges the persistent and dynamic nature of cyber threats, taking a proactive stance by eliminating inherent trust in users or devices. Trough the implementation of rigorous access controls, continuous verification processes, and resource segmentation, ZT aims to booster security measures, addressing both external attacks and potential insider risks.

This paper contributes to the exploration of ZT security paradigm by providing a comprehensive understanding of its principles and applications, it demonstrates the adaptability of ZT across various technological environments. We conducted a bibliometric and systematic review based on literature from the scopus database, this review revealed an increasing number of documents in recent years, different document types, subject areas, the keyword co-occurrence analysis, and the comparison with existing surveys related to the field of ZT.  We delved into its overarching concept, the implementation process, and its applications in various domains such as Cloud, IoT, and 5G/6G. With a specific focus on connected vehicles (CVs), the paper explores different approaches within this context. We examined CVs systems, assessed prevalent threats and cyber attacks they face, and analyzed various CVs trust models. Furthermore, we investigated different ZT models across CVs systems, including electric vehicle systems, intelligent and connected vehicles, connected and autonomous vehicles, the internet of vehicles, and other approaches in this context.

In conclusion, according to existing literature, we address primary challenges and open issues associated with proposed ZT model in CVs, aiming for clarity and comprehensiveness in understanding the potential of zero trust in enhancing automotive security. Prospective research endeavours may be oriented towards extending and applying ZT frameworks within both V2V and V2I communication paradigms. The synergistic integration of ZT methodologies with cutting-edge technological innovations has the potential to create a powerful defence mechanism against the evolving increase of cyber threats and attacks targeting connected vehicles. This merger of multifaceted approaches is expected to significantly increase safety and security protocols within contemporary transportation systems to address critical challenges posed by increasing connectivity and automation in vehicular networks. Consequently, it enhances safety and security within transportation systems.


\appendix

\label{appendix}

\section{List of abbreviations}

\begin{acronym}[MANET]\itemsep0pt
\acro {ZT}       Zero Trust
\acro {ZTA}       Zero Trust Architecture
\acro {CVs}       Connected Vehicles
\acro {EV}        Electric Vehicle 
\acro {IoV}       Internet of Vehicle
\acro {IoCV}        Internet of Connected Vehicle
\acro {PDP}        Policy Decision Point
\acro {PE}       Policy Engine
\acro {PA}        Policy Administrator
\acro {PEP}       Policy Enforcement Point
\acro {CDM}       Continuous Diagnostic and Mitigation
\acro {TA}       Trust Algorithm 
\acro {PKI}       Public Key Infrastructure
\acro {LDAP}       Lightweight Directory Access Protocol
\acro {SEIM}    Security Information and Event Management
\acro {CAN}       Controller Area Network
\acro {MFA}       Multi-Factor Authentication 
\acro {BYOD}       Bring Your Own Device 
\acro {IAM }        Identity and Access Management
\acro {GPS }        Global Positioning System
\acro {ECU }        Electronic Component Unit
\acro {LIN }        Local Interconnect Network
\acro {MOST }        Media Oriented Systems Transport
\acro {IVN }        Intra-Vehicle Network
\acro {IVC }        Inter-Vehicle Communication
\acro {ITS }        Intelligent Transportation System
\acro {VANET}        Vehicular Ad Hoc Network
\acro {MANET}        Mobile Ad Hoc Network
\acro {OBU}        On Board Unit
\acro {RSU}       Road Side Unit
\acro {DSRC}       Dedicated Short Range Communication
\acro {WAVE}       Wireless Access in Vehicular Environment
\acro {RAN}       Radio Access Network
\acro {V2V}       Vehicle to Vehicle
\acro {V2I}       Vehicle to Infrastructure
\acro {V2X}       Vehicle to Everything
\acro {V2P}       Vehicle to Pedestrian
\acro {C-V2X}       Cellular Vehicle to Everything
\acro {VCC}       Vehicular Cloud Computing
\end{acronym}

 \bibliographystyle{IEEEtran}
\bibliography{reportR}
\end{document}